# Generative AI-enhanced Sector-based Investment Portfolio Construction

Alina Voronina[1,3] · Oleksandr Romanko[2,3] · Ruiwen Cao[4,3] · Roy H. Kwon[3] · Rafael Mendoza-Arriaga[2]

**Abstract**   This paper investigates how Large Language Models (LLMs) from leading providers (OpenAI, Google, Anthropic, DeepSeek, and xAI) can be applied to quantitative sector-based portfolio construction. We use LLMs to identify investable universes of stocks within S&P 500 sector indices and evaluate how their selections perform when combined with classical portfolio optimization methods. Each model was prompted to select and weight 20 stocks per sector, and the resulting portfolios were compared with their respective sector indices across two distinct out-of-sample periods: a stable market phase (January-March 2025) and a volatile phase (April-June 2025).

Our results reveal a strong temporal dependence in LLM portfolio performance. During stable market conditions, LLM-weighted portfolios frequently outperformed sector indices on both cumulative return and risk-adjusted (Sharpe ratio) measures. However, during the volatile period, many LLM portfolios underperformed, suggesting that current models may struggle to adapt to regime shifts or high-volatility environments underrepresented in their training data. Importantly, when LLM-based stock selection is combined with traditional optimization techniques, portfolio outcomes improve in both performance and consistency.

This study contributes one of the first multi-model, cross-provider evaluations of generative AI algorithms in investment management. It highlights that while LLMs can effectively complement quantitative finance by enhancing stock selection and interpretability, their reliability remains market-dependent. The findings underscore the potential of hybrid AI-quantitative frameworks, integrating LLM reasoning with established optimization techniques, to produce more robust and adaptive investment strategies.

Alina Voronina
alina.voronina@ucu.edu.ua

Oleksandr Romanko
oleksandr.romanko@sscinc.com

Ruiwen Cao
ruiwen.cao@connect.polyu.hk

Roy H. Kwon
rkwon@mie.utoronto.ca

Rafael Mendoza-Arriaga
rafael.mendoza@sscinc.com

1 Faculty of Applied Sciences, Ukrainian Catholic University, 2a Kozelnytska Street, Lviv, Ukraine

2 SS&C Algorithmics, 200 Front Street West, suite 2500, Toronto, ON, M5V3K2, Canada

3 Department of Mechanical and Industrial Engineering, University of Toronto, 5 King's College Road, Toronto, ON, M5S3G9, Canada

4 School of Accounting & Finance, M715, 7/F, Li Ka Shing Tower, Hong Kong Polytechnic University, Hung Hom, Kowloon, Hong Kong

# 1 Introduction

Since the launch of OpenAI's ChatGPT, large language models (LLMs) have started transforming how organizations across industries use artificial intelligence. Finance is no exception. Researchers and professionals are now testing how LLMs can support tasks ranging from sentiment analysis and financial text mining to risk modeling and investment strategies. These models have also shown potential in financial advisory, trading, and portfolio management [1].

Even with their rapid progress, LLMs face important challenges when applied to finance. Much of this stems from how these systems work. Their performance depends heavily on the quality and relevance of the data they're trained on, and because they learn from patterns rather than explicit logic, it's often unclear how or why a particular output was generated. This "black-box" nature raises understandable concerns about accuracy, reliability, and transparency. As Zaremba et al. note [2], the quality of data used to train Natural Language Processing (NLP) models directly affects the precision and stability of their results. For that reason, models designed with financial data in mind are often better suited to handle market-specific tasks than general-purpose ones.

Another issue lies in consistency. Generative AI models sometimes produce different or even contradictory answers when given similar prompts over time. They're also sensitive to imbalances or biases in their training data, which can influence the fairness and reliability of their outputs [3].

Interpretability is another critical challenge. Because LLMs don't provide a clear reasoning trail, users can't always explain or verify their recommendations. Improving this aspect is essential if AI systems are to earn trust in financial decision-making. As Zaremba et al. emphasize [2], better explainability will be a key step toward using these tools confidently in risk-sensitive areas such as investment management.

Despite these hurdles, the potential of LLMs in finance continues to grow. Romanko et al. [4], for example, explored how ChatGPT (GPT-4) could be used to construct diversified portfolios of S&P 500 stocks and even attempt to outperform the index. Building on this idea, our research expands the scope to include a wider set of AI models and multiple market indices, allowing for broader and more robust comparisons.

In this study, we analyze several leading LLMs, OpenAI's models, Google's Gemini, Anthropic's Claude, DeepSeek's models, and xAI's Grok, to evaluate how well they perform on a common set of financial tasks. The key question we ask is whether general-purpose LLMs can effectively select stocks and assign portfolio weights that outperform traditional benchmarks. Our analysis is based on the eleven S&P 500 sector indices: energy, materials, industrials, consumer discretionary, consumer staples, health care, financials, information technology, communication services, utilities, and real estate.

This work builds on earlier single-model research by expanding it to a comparative study of twelve different LLMs. We also move from an index-level to a sector-level framework, which allows us to see how model performance varies across different areas of the market. In addition, the effective construction of sector index portfolios underlies ETF sector indices which are useful for practical investment purposes such as gaining exposure to a particular market segmens. The goal is not only to measure performance but also to address recurring issues such as explainability and reliability.

Importantly, AI should not replace human judgment. As highlighted by [5], it should serve as one part of a broader decision-making process that combines data-driven insights with human experience. Our goal is to determine whether LLMs can act as valuable tools that enhance, rather than replace, traditional portfolio management methods.

Our research makes several contributions:

1. We expand AI-driven portfolio construction into a multi-model framework, comparing twelve of the most advanced LLMs available in 2025.
2. We perform the analysis at the sector level to understand how model performance varies across industries.
3. We use two out-of-sample testing periods, a stable one (January-March 2025) and a volatile one (April-June 2025), to see how results change under different market conditions.
4. Finally, we combine LLM-based stock selection with traditional mean–variance optimization to test whether blending generative reasoning with quantitative techniques can improve portfolio stability and returns.

The findings are encouraging but nuanced. During stable market periods, LLMs can build portfolios that outperform sector benchmarks, showing stronger cumulative returns and Sharpe ratios. However, their performance drops during volatile conditions, suggesting that market stability and the presence of similar data in training play major roles in success. Combining LLM insights with optimization models tends to produce more consistent results, pointing toward hybrid systems where AI and quantitative tools work together as the most promising path forward.

Across models, we observed similar behavioral patterns, though results varied by sector. The Energy, Financials, and Information Technology sectors generally performed best, while Consumer Staples tended to lag. Statistical analysis showed that these differences were linked less to diversification and more to volatility, aligning with classical risk–return dynamics.

Overall, LLMs demonstrated strong risk-adjusted performance in stable markets, supporting their value as analytical partners for investment research and stock selection. When conditions become turbulent, however, their limitations are more apparent. In practice, this means LLMs work best as supportive tools, generating ideas, highlighting opportunities, and providing reasoning that complements human expertise and traditional quantitative methods.

In summary, this study contributes to the growing conversation about how AI can be responsibly integrated into finance. It offers one of the first large-scale, comparative analyses of LLMs in quantitative investment management, identifying the situations where they excel and where they fall short. Ultimately, the findings point to a future where hybrid systems, blending human insight, quantitative rigor, and AI-driven reasoning, create more transparent, stable, and effective financial decision-making frameworks.

## 2 Related Work

Recent studies have begun exploring how large language models (LLMs) can support investment decision-making and portfolio management. Biswas et al. [5] and Ko et al. [6] highlight the growing potential of generative AI as a transformative tool for investors. These models can assist in analyzing data, generating investment insights, and even serving as educational resources for individuals learning about portfolio construction [6]. However, their reliability and consistency remain under active scrutiny. One major concern is "hallucinations", a phenomenon where models produce inaccurate or fabricated information, which can undermine the credibility and authenticity of their outputs.

Beyond hallucinations, other challenges also affect the use of generative AI in finance. These include dependence on data quality, limited transparency, risks of data manipulation, security issues, and the fast-changing nature of financial markets and regulations. Overreliance on AI-generated recommendations can also lead to poor investment choices that lack sound analytical reasoning [5].

The development of domain-specific models has emerged as one way to address these challenges. BloombergGPT [7] is a leading example of an LLM designed specifically for financial tasks. Trained on a large, finance-focused dataset, BloombergGPT demonstrates exceptional performance on financial benchmarks while maintaining strong results on more general tasks [8]. The authors compared its performance across two dimensions, general-purpose tasks and finance-specific tasks, and found that while it performed competitively overall, it clearly outperformed other models on financial applications. This highlights the benefits of tailoring LLMs to specialized domains where contextual understanding and data precision are critical.

In addition to BloombergGPT, there are GPT-4–based systems built specifically for stock analysis and selection. One such model, MarketSenseAI, integrates diverse data sources, including company updates, market trends, and macroeconomic indicators, to generate investment recommendations [9]. Despite its innovative design, MarketSenseAI has certain limitations. The model's predictions were tested only over a short time horizon, restricting its ability to adapt to different market conditions or sustain performance across economic cycles [9].

While models like BloombergGPT and MarketSenseAI demonstrate the enormous potential of LLMs in finance, they also underline ongoing limitations around interpretability, explainability, and transparency. These issues remain central to whether investors and institutions will trust AI-driven tools in high-stakes financial decision-making. The challenges apply equally to domain-specific and general-purpose models, underscoring the need for responsible integration of AI into financial workflows. Addressing these concerns is essential for realizing the full promise of AI while maintaining fairness, accountability, and human oversight.

Romanko et al. [4] provide one of the most detailed explorations of generative AI applied directly to portfolio selection. In their study, the authors used OpenAI's GPT-4 to construct portfolios of 15, 30, and 45 stocks drawn from the S&P 500 index, then compared the performance of these portfolios against the S&P 500 itself and optimized versions based on different criteria.

Their approach relied on prompting GPT-4 to select $n$ stocks (where $n$ = 15, 30, or 45) with the goal of outperforming the index. Because generative models may produce different outputs even when given identical prompts, the authors repeated the same request several times. They then identified the most frequently selected stocks across these runs to build more stable portfolios.

Five portfolios were constructed for comparison:

- a GPT-weighted portfolio, where ChatGPT both selected and assigned weights to stocks;
- a GPT equally weighted portfolio, where selected stocks were equally weighted;
- a minimum variance portfolio,
- a maximum expected return portfolio, and
- a maximum Sharpe ratio portfolio, all built using the same stock universe suggested by GPT-4.

These portfolios were benchmarked not only against the S&P 500 but also against the Dow Jones Industrial Average and the Nasdaq Composite indices. To further assess real-world relevance, the results were also compared to portfolios managed by major asset management firms such as Vanguard, Fidelity, and BlackRock [4].

The study concluded that ChatGPT performs surprisingly well as a stock selector and that its performance improves even further when paired with established optimization frameworks. This finding suggests that the most effective use of generative AI in portfolio management may lie in hybrid approaches – where AI-generated insights complement traditional quantitative methods and human expertise.

## 3 Stock Universe Construction and Data Sources

The S&P 500 Sector Indices are subindices of the S&P 500 that group companies according to their primary lines of business using the Global Industry Classification Standard (GICS) [10]. Each S&P 500 company belongs to exactly one sector index, ensuring a clear and consistent classification hierarchy across the broader benchmark.

There are eleven sectors represented in the S&P 500: Energy, Materials, Industrials, Consumer Discretionary, Consumer Staples, Health Care, Financials, Information Technology, Communication Services, Utilities, and Real Estate. Each sector is tracked through its own index, with the official names and tickers summarized in Table 1.

Because S&P Dow Jones Indices does not publicly release the constituent lists for each sector index, the sector universes used in this study were reconstructed following the official S&P U.S. Indices Methodology [10]. The goal was to replicate each sector index as accurately as possible using publicly available data sources.

To verify that the reconstructed sector lists match the official indices, we performed several consistency checks:

- Constituent counts were cross-checked with the official numbers published on the S&P Global website [11];

- The full S&P 500 constituent list from Wikipedia [12] was used to identify which companies belong to each sector according to the GICS classification.
- For an additional layer of validation, the resulting lists were compared with the sector compositions available on TradingView [13], ensuring close alignment between reconstructed and market-referenced sector compositions.

This validation confirmed that the number of companies assigned to each sector matched the official counts and that all companies included were legitimate S&P 500 constituents. This process provided a consistent and transparent foundation for further analysis.

| S&P-500 sector | Ticker |
|---|---|
| Energy | S&P 500-10 |
| Materials | S&P 500-15 |
| Industrials | S&P 500-20 |
| Consumer Discretionary | S&P 500-25 |
| Consumer Staples | S&P 500-30 |
| Health Care | S&P 500-35 |
| Financials | S&P 500-40 |
| Information Technology | S&P 500-45 |
| Communication Services | S&P 500-50 |
| Utilities | S&P 500-55 |
| Real Estate | S&P 500-60 |

Table 1: S&P 500 sector names and their corresponding tickers

Because most of the large language models (LLMs) used in this study have training cut-off dates before 2025, we divided the analysis into one in-sample period and two out-of-sample periods:

- The in-sample period spans the five years preceding January 2025.
- The first out-of-sample period runs from January to March 2025, a relatively stable market phase.
- The second out-of-sample period, from April to June 2025, coincides with high volatility period following significant market disruptions.

This structure enables a clear assessment of how well LLM-generated portfolios perform both in normal and turbulent conditions, and how effectively they generalize beyond their training data.

Market data were obtained primarily from Yahoo Finance, which provided adjusted closing prices for both individual company stocks and the S&P 500 sector indices. These prices form the basis for calculating portfolio returns, risk measures, and optimization inputs throughout the study.

While most sector indices follow the standard ticker conventions shown in Table 1, it should be noted that on Yahoo Finance the Energy sector is listed under the ticker *GSPE*.

Through this structured data assembly process, combining official S&P Global methodology, publicly available listings, and third-party validation, we established a transparent and verifiable dataset of sector constituents. This dataset serves as the foundation for constructing and evaluating LLM-generated portfolios across all eleven S&P 500 sectors.

## 4 Methodology

This section describes the methodology used to evaluate how large language models (LLMs) perform in sector-based portfolio construction. Twelve models from major AI developers were tested across the eleven S&P 500 sectors, each prompted to select and weight stocks using a consistent structure. The selected stocks were verified against official sector constituents, and invalid entries were adjusted to ensure full portfolio allocation. The resulting portfolios were then evaluated using the mean–variance optimization framework to compare LLM-generated results with those produced through traditional quantitative methods. For each sector and model, five portfolios were analyzed: LLM-weighted, equally weighted, minimum variance, maximum return, and maximum Sharpe ratio, forming the basis for assessing whether AI-driven reasoning can enhance classical investment strategies.

### 4.1 Model Selection

This study evaluates a diverse set of twelve large language models (LLMs) developed by five leading AI companies: OpenAI, Google, Anthropic, DeepSeek, and xAI. Each model represents a unique architecture and training procedure, offering an opportunity to compare how design differences affect performance in financial portfolio construction tasks.

Because model capabilities evolve rapidly, we include multiple versions from several providers to capture both cross-developer and cross-generation variations. The following subsections summarize the specific models used and their relevant technical details.

OpenAI's models serve as a benchmark in this research, given their widespread adoption and continuous improvement in reasoning and data interpretation. Four of the company's most recently released models are included:

- GPT-4o, released in May 2024 and trained on data through September 2023, is known for combining speed with robust reasoning performance.
- GPT-4.1, released in April 2025 (gpt-4.1-2025-04-14), incorporates data up to May 2024 and demonstrates more advanced analytical accuracy and reasoning stability.
- o4-mini, introduced alongside GPT-4.1, is a smaller reasoning-optimized model trained on the same data window (to May 2024) and designed for efficient performance in constrained computational settings [14].

- GPT-5, OpenAI's newest flagship model, officially launched on August 7, 2025, with a knowledge cut-off of September 30, 2024. According to OpenAI and independent reports, GPT-5 achieves near PhD-level competency across reasoning, writing, and coding domains [15].

Including multiple OpenAI models allows us to examine how incremental training-data expansion and architectural refinements influence financial reasoning and decision quality.

From Google's Gemini family, we include Gemini 2.5 Pro, released in March 2025, with a knowledge cut-off of January 2025 [16]. Gemini is designed for advanced multimodal reasoning, integrating text, data, and contextual analysis. Its inclusion enables direct comparison with OpenAI's models that share similar data horizons but differ substantially in architecture and learning approach.

Three Anthropic models are examined to capture the evolution of the Claude series toward higher-order reasoning and interpretability:

- Claude Sonnet 3.7, released in early 2025 and trained on data through November 2024, improves upon earlier versions by extending context windows and deep-reasoning abilities.
- Claude Sonnet 4 and Claude Opus 4, both released in May 2025, are the most advanced in the family, designed to handle complex analytical, writing, and research tasks with greater stability and transparency [17].

These versions enable evaluation of how model maturity within a single architecture affects the quality of financial analysis and portfolio construction.

Two models from xAI are included: Grok 3 and Grok 3 Mini, both released in February 2025 [18]:

- Grok 3 is a broad conversational model with strong domain coverage in areas such as finance, law, and science.
- Grok 3 Mini is optimized for deliberate reasoning, designed to "think before responding".

Their inclusion allows testing whether smaller, reasoning-focused models offer advantages in consistency and explainability compared with larger, general-purpose variants.

Finally, the study incorporates two models from DeepSeek:

- DeepSeek-V3 (deepseek-chat) and,
- DeepSeek-R1 (deepseek-reasoner).

Both were trained on data through July 2024 and represent different design philosophies, one tuned for conversational precision and the other for structured analytical reasoning. This pairing provides insight into how specialized reasoning mechanisms affect portfolio-selection accuracy and stability.

Some models' training periods overlap partially with the study's out-of-sample evaluation windows. For instance, the Claude 4 models were trained on data extending into early 2025, which may give them limited exposure to the same market conditions used for testing.
This overlap is acknowledged when interpreting comparative performance to ensure fair analysis.

Altogether, the research evaluates twelve LLMs, four from OpenAI, three from Anthropic, two each from DeepSeek and xAI, and one from Google. By applying all models to an identical financial-portfolio-construction framework, we can systematically assess:

- how differing architectures and reasoning styles influence stock-selection quality,
- the relative impact of recent training data on performance, and
- which model families show the most robustness across market regimes.

This comprehensive list forms the foundation for a rich multi-model, cross-provider evaluations of LLMs in quantitative investment management.

### 4.2. Selection of Trading Universe of Stocks

To construct sector-based portfolios, we used application programming interfaces (APIs) from multiple LLM providers to obtain each model's response to a standardized prompt requesting the selection of stocks from a specific S&P 500 sector index. The goal was to ensure consistency in how prompts were phrased while allowing each model to apply its internal reasoning when identifying potentially outperforming securities.

Each prompt was tailored to the sector being analyzed. For example, the prompt used for the S&P 500-20 Industrials sector index appeared as follows:

> *Using a range of investing principles taken from leading funds, create a theoretical fund comprising of at least 20 stocks (mention their tickers) from the S&P500-20 industrials sector with the goal to outperform the S&P 500-20 industrials sector (S&P 500 sector index).*

The variables "tickers" and "sector index" were dynamically replaced depending on the sector being processed. This formulation ensured that each model's response produced a realistic sector-constrained portfolio along with its reasoning for choosing particular securities. Because generative models often return variable outputs, each prompt was run 10 times per sector to capture a representative range of responses.

After receiving a portfolio composition, a second prompt was issued to extract only the stock tickers:

> *Extract only the ticker symbols of the stocks comprising the fund from the previous response. In your response to this prompt, list only the ticker symbols separated by spaces.*

Figure 1 and Figure 2 illustrate examples of the resulting stock selections for the Information Technology and Industrials sectors generated by the Claude Sonnet 3.7 model. The highlighted (green) tickers represent the 20 most frequently mentioned valid companies within each sector.

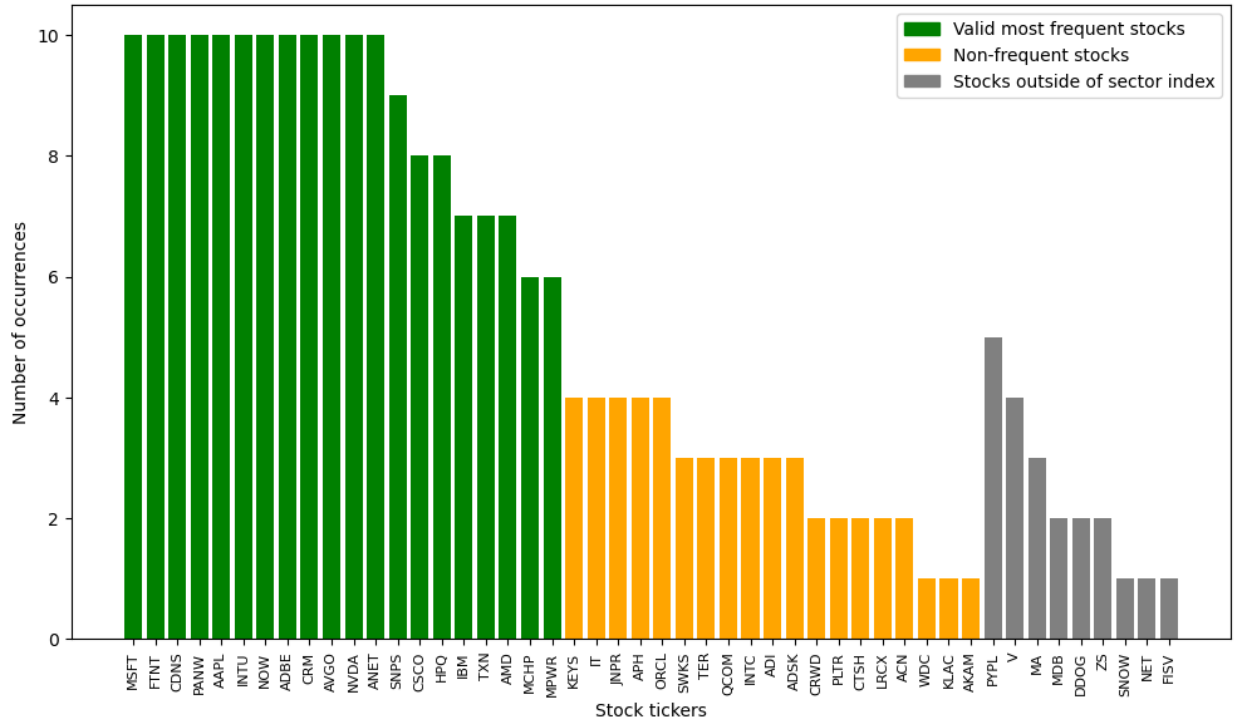

Figure 1: Selected stocks in Information Technology sector for Claude Sonnet 3.7 model

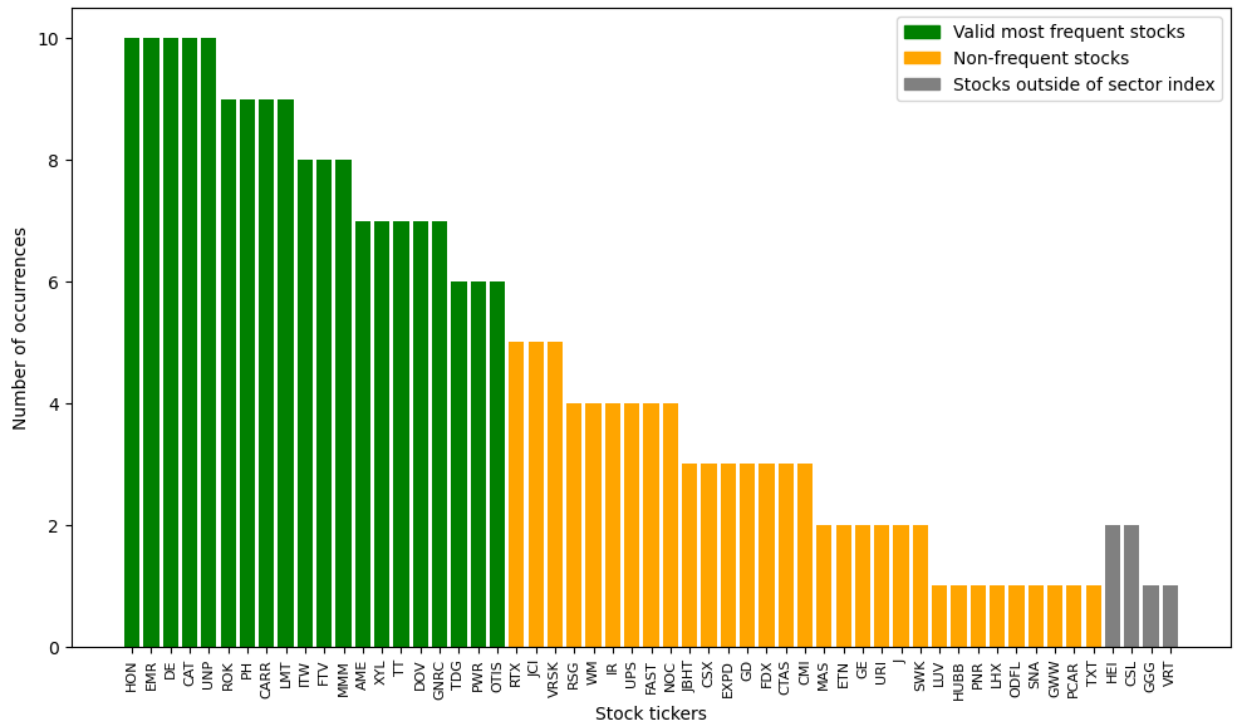

Figure 2: Selected stocks in Industrials sector for Claude Sonnet 3.7 model

To ensure data integrity, each stock proposed by the models was validated against the official list of S&P 500 sector constituents described in Section 3. Companies not present in the appropriate sector index were removed. In addition, stocks for which Yahoo Finance did not provide a complete historical price series were excluded to maintain consistency in return calculations.

Once valid sector constituents were confirmed, a follow-up prompt requested each LLM to assign hypothetical portfolio weights to its selected stocks. The prompt for the S&P 500-20 Industrials sector, for example, was:

> *Assume you're designing a theoretical model portfolio from these S&P 500-20 Industrials stocks: AB CD EFG. Provide a hypothetical example of how you might distribute the weightage of these stocks (normalized i.e weights should add up to 1.00) in the portfolio to potentially outperform the S&P 500-20 Industrials index. Also mention the underlying strategy or logic which you used to assign these weights.*

Here, *"AB CD EFG"* was replaced with the ticker list produced by the previous step. Each model's response typically contained both the numeric weights and its underlying rationale for allocation decisions. Because responses varied between runs, this weight-assignment step was executed five times for each sector. The final portfolio weights were then determined by taking the average normalized weight across all iterations for each stock.

To convert those responses into machine-readable format, a fourth prompt was used:

> *Extract tickers of stocks and corresponding weights as a single comma "," separated string, with the weights expressed as floats: {output}. Provide a list of type TICKER: weight, no extra symbols used, and no extra text or explanation. Example output: "AAPL: 0.2, MSFT: 0.3, GOOGL: 0.5"*

The placeholder *{output}* represents the text returned by the model in response to prompt 3.

This approach builds directly on the methodology of Romanko et al. [4], who used a similar prompt to generate portfolios from the full S&P 500 index. In the present study, the prompt structure was refined to better align with sector-level analysis and to improve the consistency of model outputs. The following additions were made:

- concise instructions such as *"no extra text or explanation"* to enforce clean output formatting; and
- an explicit example of expected response structure, e.g., *"AAPL: 0.2, MSFT: 0.3, GOOGL: 0.5."*

These refinements helped models produce structured results suitable for downstream quantitative processing.

Executing these four prompts for each of the eleven sectors across all twelve LLMs produced 132 unique sector portfolios. Each portfolio corresponded to a specific model-sector combination. For each of these, we also created a secondary equally weighted version of the same portfolio, assigning weights of $1/n$ (where $n$ = number of selected stocks). This resulted in an additional 132 portfolios, bringing the total to 264 across both weighting schemes:

- Model-selected, model-weighted portfolios: LLM both selects and weights stocks.

- Model-selected, equally weighted portfolios: LLM selects stocks, but all receive equal weighting.

*This structured, prompt-driven process ensured a consistent and transparent way to compare how different LLMs approach stock selection and portfolio construction. By leveraging clearly defined prompts and repeated sampling, we minimized random variability and established a robust foundation for the subsequent optimization and performance-evaluation stages described in the following section.*

### 4.3 Mean-Variance Portfolio Optimization Model

Once the LLM-generated portfolios were established, they were evaluated and optimized using the mean–variance portfolio optimization (MVPO) framework, originally introduced by Harry Markowitz [19]. This classical quantitative approach provides a rigorous and mathematically grounded method for balancing expected return against portfolio risk (measured by variance) and remains a foundational concept in modern portfolio theory.

In this research, the optimization process serves two complementary objectives:

- To compare the performance of the raw LLM-constructed portfolios (as generated in Section [4.2](#)) with portfolios optimized using traditional quantitative methods.
- To assess whether combining LLM-driven stock selection with quantitative optimization can yield superior and more stable performance than either method alone.

The optimization seeks an allocation vector $w$ that either minimizes portfolio variance for a given expected return or maximizes expected return for a given level of risk. The optimization problem can be expressed through the following objective function, subject to a set of constraints:

$$\begin{aligned} \min_{w} \quad & w^T \cdot Q \cdot w \\ \text{s.t.} \quad & \mu^T \cdot w \geq \varepsilon \\ & \Sigma_{i=1}^{n} w_i = 1 \\ & l \leq w \leq u \end{aligned} \quad (1)$$

where:

- $\mu$ is the vector of expected returns,
- $Q$ is the covariance matrix of asset returns,
- $\varepsilon$ is the target expected return,
- $l$ and $u$ are the lower and upper bounds for individual asset weights.

The mean-variance efficient frontier is obtained by varying the hyperparameter $\varepsilon$, which defines the target level of expected return. Adjusting $\varepsilon$ allows for the generation of multiple efficient portfolios along the frontier, representing optimal trade-offs between risk and reward.

For each model and each sector, three additional optimized portfolios were computed using standard quantitative finance algorithms. These portfolios correspond to points on the efficient frontier.

Romanko et al. [4] employed a similar approach to evaluate the risk-reward profiles of portfolios built from different trading universes consisting of 15, 30, and 45 S&P 500 stocks. In contrast, our research uses sector-level universes of 20 stocks selected by the LLMs from S&P 500 sector indices. This portfolio size was chosen to balance diversification with the practical variation in sector sizes, ensuring representative coverage without excessive dilution.

The mean-variance efficient frontier is computed for each LLM-defined universe, and three optimized portfolios are then selected [19]:

- Minimum Variance Portfolio (MVP): minimizes portfolio variance;
- Maximum Expected Return Portfolio (MERP): maximizes expected return;
- Maximum Sharpe Ratio Portfolio (MSRP): maximizes the Sharpe ratio by balancing risk and reward.

As described in Romanko et al. [4], the optimization problem includes constraints ensuring that:

$$\begin{array}{c} \Sigma_{i=1}^{n} w_i = 1 \\ l \leq \boldsymbol{w} \leq u \end{array} \quad (2)$$

where $n$ is the total number of assets in a given universe, and $\boldsymbol{w}$ represents the vector of portfolio weights. The second constraint ($l \leq \boldsymbol{w} \leq u$) enforces realistic boundaries on asset weights, ensuring that each stock maintains a non-zero position, thus preserving a consistent portfolio size across all constructed universes.

Following Romanko et al. [4], the lower bound $l$ is set to approximately half of the equally weighted portfolio weight, i.e., $\frac{1}{2}\widehat{w}$, while the upper bound $u$ is set to roughly twice that value $w$, i.e., $2\widehat{w}$, where $\widehat{w}$ is $1/n$. For a portfolio of 20 stocks, these bounds translate to:

$$0.025 \leq \boldsymbol{w} \leq 0.10$$

This constraint framework ensures that all portfolios remain diversified yet flexible, preventing over-concentration in individual positions while maintaining stability across optimization runs.

The optimization problem, as defined by Equation (1), is solved iteratively to compute the efficient frontier for each LLM-derived trading universe.

Figures 3 and 4 illustrate examples of these bounded efficient frontiers (solid lines) for the Industrials and Information Technology sectors using the GPT-4o model (based on in-sample data from the five years preceding January 2025).

Each figure highlights the three optimized portfolios on the efficient frontier, Minimum Variance, Maximum Expected Return, and Maximum Sharpe Ratio, alongside the two LLM-generated portfolios (LLM-weighted and equally weighted). For comparison, the dashed line represents the efficient frontier obtained under a simpler constraint ($w \geq 0$) instead of the bounded constraints $l \leq \boldsymbol{w} \leq u$ above. Individual stocks selected by the LLMs are plotted as discrete points to visualize their relative positions in terms of return and risk.

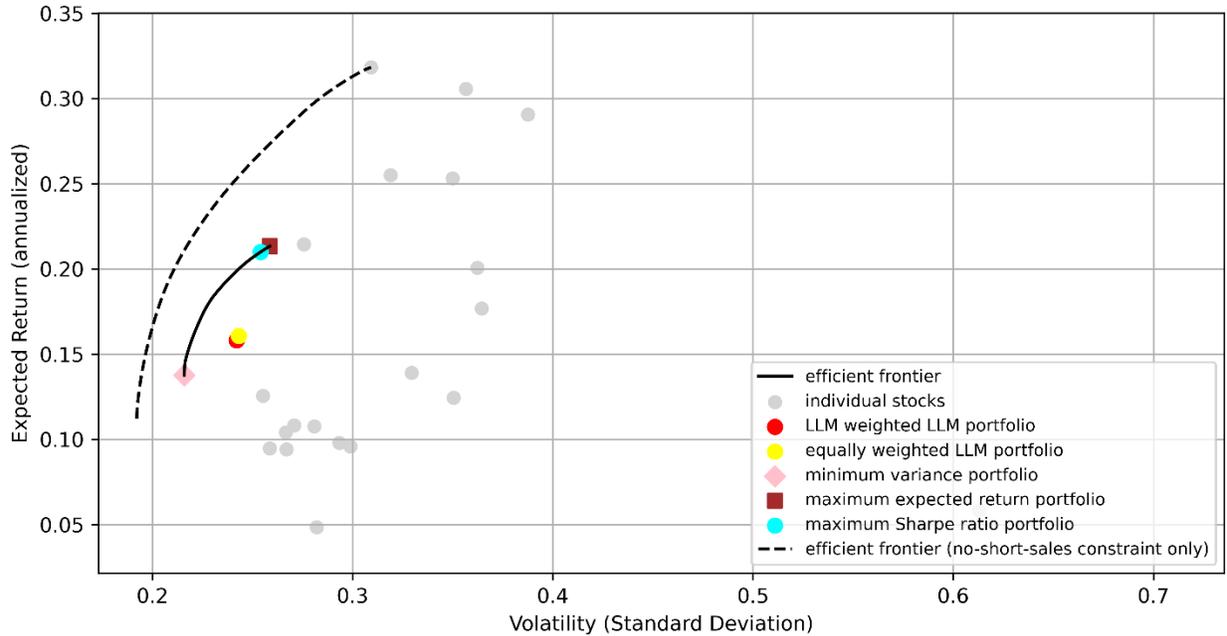

Figure 3: Efficient frontier for the portfolio in Industrials sector for GPT-4o model (in-sample: 5 years prior to January 2025)

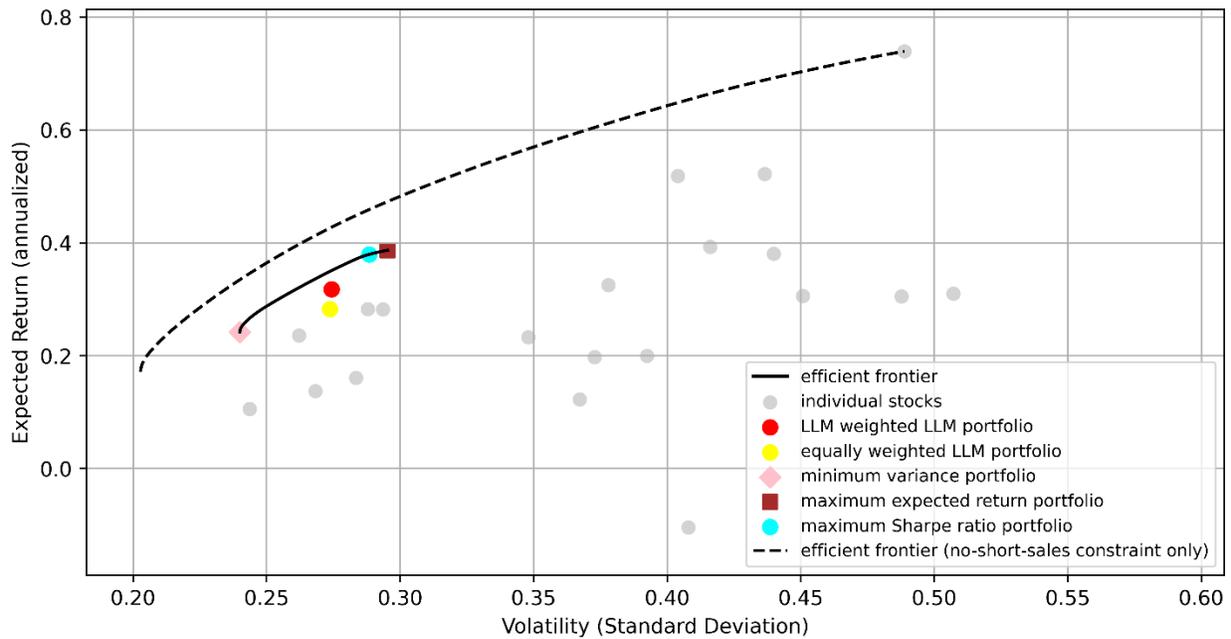

Figure 4: Efficient frontier for the portfolio in Information Technology sector for GPT-4o model (in-sample: 5 years prior to January 2025)

At the conclusion of this step, five portfolios are derived for each LLM-generated sector universe:

- LLM-weighted portfolio (stocks and weights generated by the model);
- Equally weighted portfolio (LLM-selected stocks with equal weights);
- Minimum variance portfolio (LLM-selected stocks are used as trading universe);

- Maximum expected return portfolio (LLM-selected stocks are used as trading universe);
- Maximum Sharpe ratio portfolio (LLM-selected stocks are used as trading universe).

Given that 12 models each produce portfolios for 11 S&P 500 sectors, a total of 132 unique model–sector combinations (universes) were analyzed. This process resulted in the creation of 660 portfolios overall, forming the basis for subsequent comparative performance evaluation.

Table 2 summarizes the computational framework used across portfolio sizes, weighting approaches, and optimization methods. Additionally, experiments with smaller universes of 15 stocks were conducted to analyze how the number of assets influences portfolio behavior and performance outcomes.

| Weight assignment (portfolio) model | Portfolio size | |
|---|---|---|
| | 15 assets | 20 assets |
| **LLM weighted** | LLM-weighted portfolio | LLM-weighted portfolio |
| **LLM universe of stocks, equal weights** | Equally-weighted portfolio | Equally-weighted portfolio |
| **Continuous optimization (efficient frontier) using LLM universe of stocks** | Min Var, Max Ret, Max Sharpe portfolios | Min Var, Max Ret, Max Sharpe portfolios |

Table 2: Computational framework for each sector index (applied to all 12 LLMs)

This integrated framework, combining generative AI reasoning with traditional mean-variance optimization, provides a structured and replicable approach to evaluating whether LLM-driven stock selection can enhance quantitative portfolio construction.

## 5 Computational Experiments

This section presents the results and analysis of the study, highlighting how large language models perform in constructing and optimizing sector-based investment portfolios. The findings show that LLMs can identify well-performing stocks and generate competitive portfolios, particularly under stable market conditions, where they often outperform their respective sector benchmarks. However, performance declines during volatile periods, indicating sensitivity to changing market dynamics and training data limitations. The analysis also reveals that combining LLM-driven stock selection with traditional mean-variance optimization improves portfolio stability and overall risk-adjusted returns. Among sectors, Information Technology, Financials, and Energy tend to yield the strongest results, while Consumer Staples consistently underperform. Across all models, results confirm that LLMs are most effective when used as analytical tools that complement, rather than replace, quantitative methods and human expertise in investment decision-making.

### 5.1 Comparing LLM Reasonings

We began by examining the reasoning patterns expressed by each large language model (LLM) during the stock-selection and weight-assignment stages. Table 3 summarizes these rationales, with each "+" symbol indicating that the model (column) employed a specific investment reasoning (row). By analyzing both shared and distinctive elements across models, we identified common investment philosophies as well as architecture-specific behavioral nuances.

Across all twelve models, several foundational principles consistently emerged. The most prevalent were the preference for high-quality market leaders, a focus on high-growth potential, and a strong emphasis on risk management. Together, these form the backbone of many classical portfolio-construction frameworks. Many models also demonstrated an explicit bias toward underweighting weaker or financially challenged firms, which reflects a quality- or momentum-driven investment orientation.

Diversification, particularly across sub-sectors, was another widely shared theme. Some models, such as GPT-4o and o4-mini, extended this reasoning further by recommending equal weighting within similar sub-sectors, suggesting an awareness of concentration risk even within single-sector portfolios.

Nearly all models also incorporated defensive stock selection, particularly in the context of heightened macroeconomic uncertainty. Despite these overarching similarities, several models displayed more distinctive approaches that revealed deeper strategic reasoning. For instance, the Claude, DeepSeek, and Grok families explicitly incorporated geographic diversification, indicating that their selection logic extended beyond sector fundamentals to include global macro-level considerations potentially mitigating exposure to regional or geopolitical risks.

Certain models, notably GPT-4.1 and Gemini, applied a core-satellite framework, maintaining a stable core of high-conviction holdings complemented by smaller thematic or higher-risk "satellite" positions aimed at generating excess alpha. A related but more concentrated approach (the concentrated-alpha strategy) appeared in models such as Claude Opus 4 and DeepSeek-V3, where the focus was on overweighting top performers to drive outperformance.

Many of the models (including GPT-4o, GPT-5, Claude Sonnet 3.7, Claude Opus 4, and Gemini) explicitly referenced the barbell strategy, balancing aggressive growth opportunities with stable or defensive assets. This demonstrates awareness of the classical risk-return trade-off and suggests that these models attempt to emulate professional portfolio-management logic by pairing volatility-prone assets with more stable holdings.

Other nuanced approaches also emerged. Several models selectively identified value or turnaround opportunities, stocks deemed temporarily undervalued but with improving fundamentals, while Claude Sonnet 4 uniquely implemented a growth-at-a-reasonable-price (GARP) framework, combining growth potential with valuation discipline. This behavior indicates a more balanced investment philosophy that avoids overexposure to overvalued momentum stocks.

| Reason/LLM Model | GPT-4o | GPT-4.1 | o4-mini | GPT-5 | Claude Sonnet 3.7 | Claude Sonnet 4 | Claude Opus 4 | DeepSeek-V3 | DeepSeek-R1 | Gemini 2.5 Pro | Grok 3 | Grok 3 Mini |
|---|---|---|---|---|---|---|---|---|---|---|---|---|
| High-Quality Market Leaders | + | + | + | + | + | + | + | + | + | + | + | + |
| High-Growth Potential | + | + | + | + | + | + | + | + | + | + | + | + |
| Risk Management | + | + | + | + | + | + | + | + | + | + | + | + |
| Barbell Strategy | + |  |  | + |  |  | + |  |  | + |  |  |
| Factor-Based Composite Scoring |  |  | + |  |  |  |  |  | + |  |  | + |
| Underweight Weak | + | + | + | + | + | + | + | + | + | + | + | + |
| Valuation Buffer |  |  | + |  |  |  |  |  |  |  |  |  |
| Turnaround |  |  | + |  |  | + | + | + | + | + | + |  |
| Speculative Plays | + |  |  |  |  |  |  | + |  | + |  |  |
| Concentrated Alpha |  | + |  |  | + |  |  |  |  | + | + | + |
| Diversification Across Sub-Sectors | + | + | + | + | + | + | + | + | + | + | + | + |
| Equal Weighting for Sub-Sector Players | + |  |  |  |  |  |  |  |  |  | + | + |
| Core-Satellite |  | + |  |  |  |  |  |  |  | + |  |  |
| Geographic Diversification |  |  |  |  | + | + | + | + | + |  | + | + |
| Thematic Exposure | + |  |  |  |  |  |  |  |  |  |  |  |
| Defensive Plays | + | + | + | + | + | + | + | + | + | + | + | + |
| Growth at Reasonable Price |  |  |  |  |  |  | + |  |  |  |  |  |
| Competitive Moats |  |  | + | + | + | + | + | + | + | + |  |  |

Table 3: Reasonings of LLMs for stock selection and weighting when constructing a portfolio

Interestingly, the three reasoning-optimized models (o4-mini, DeepSeek-R1, and Grok 3 Mini) adopted an explicitly quantitative approach through factor-based composite scoring. These models ranked stocks according to a combination of momentum, earnings revisions, revenue growth, and value metrics before assigning weights based on the aggregate factor scores. This structured, rules-based reasoning contrasts with the more narrative, qualitative rationales offered by other models.

Overall, while all models converge on established investment principles (quality, growth, diversification, and risk control) they differ meaningfully in how these ideas are operationalized. The variations appear to stem from architectural design, training data, and reasoning orientation. Some models favor heuristic reasoning and narrative justification, whereas others employ factor-based or composite scoring frameworks resembling quantitative screening models.

Finally, it is worth noting that, although the wording of explanations varied, many models expressed conceptually similar ideas under different terms. Table 3 reflects literal phrases extracted from model outputs, illustrating both the semantic diversity and the underlying conceptual alignment across LLMs.

**5.2 Portfolio Performance Comparison based on Out-of-Sample Cumulative Returns**

For every portfolio, we evaluated out-of-sample cumulative returns over two distinct periods, January to March 2025 and April to June 2025, and compared them to the returns of their corresponding S&P 500 sector indices. In total, this produced 22 cumulative return plots across the eleven sectors and two time intervals. Figures 5-10 illustrate representative results for three sectors (Consumer Staples, Information Technology, and Utilities) showing the performance of each sector index alongside the five portfolios created by each of the twelve LLMs.

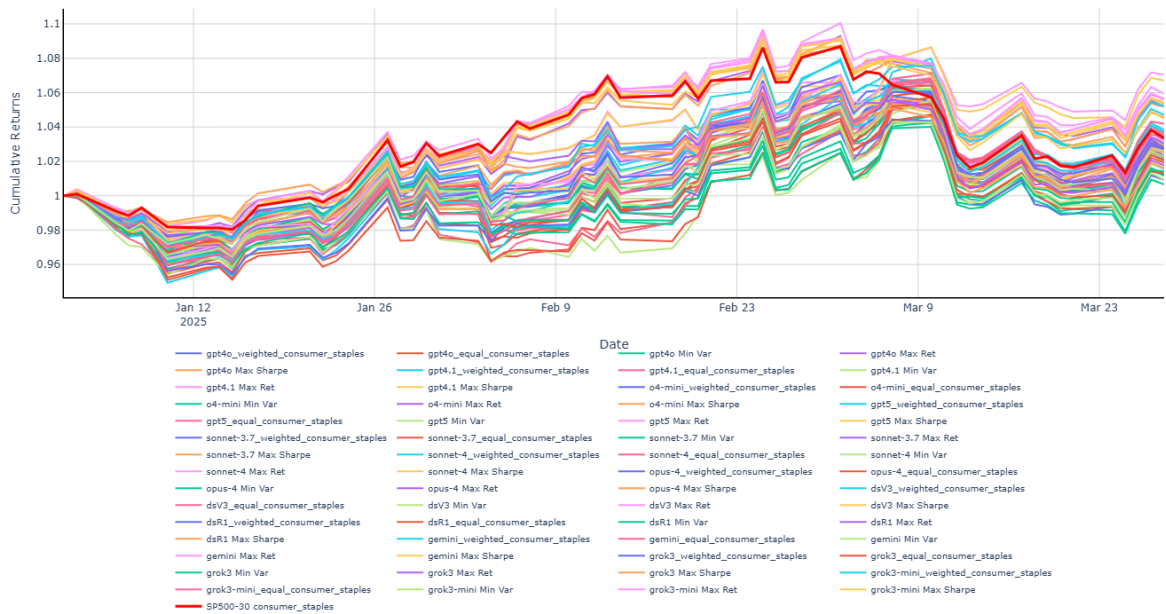

Figure 5: Out-of-sample cumulative returns of portfolios composed by LLMs for "Consumer Staples" sector (January - March 2025)

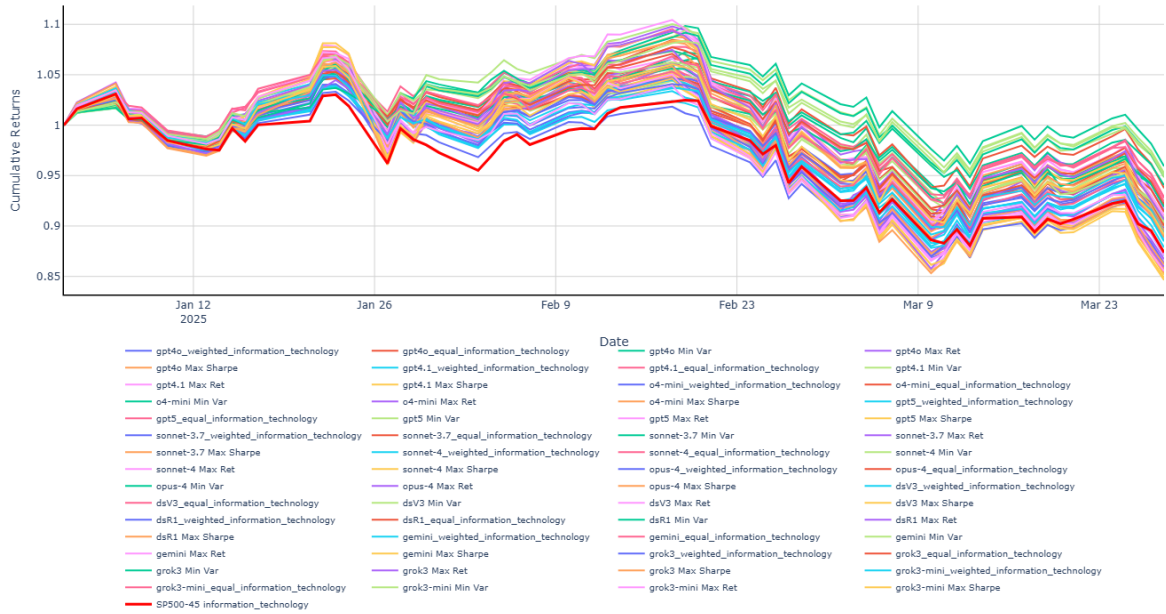

Figure 6: Out-of-sample cumulative returns of portfolios composed by LLMs for "Information Technology" sector (January - March 2025)

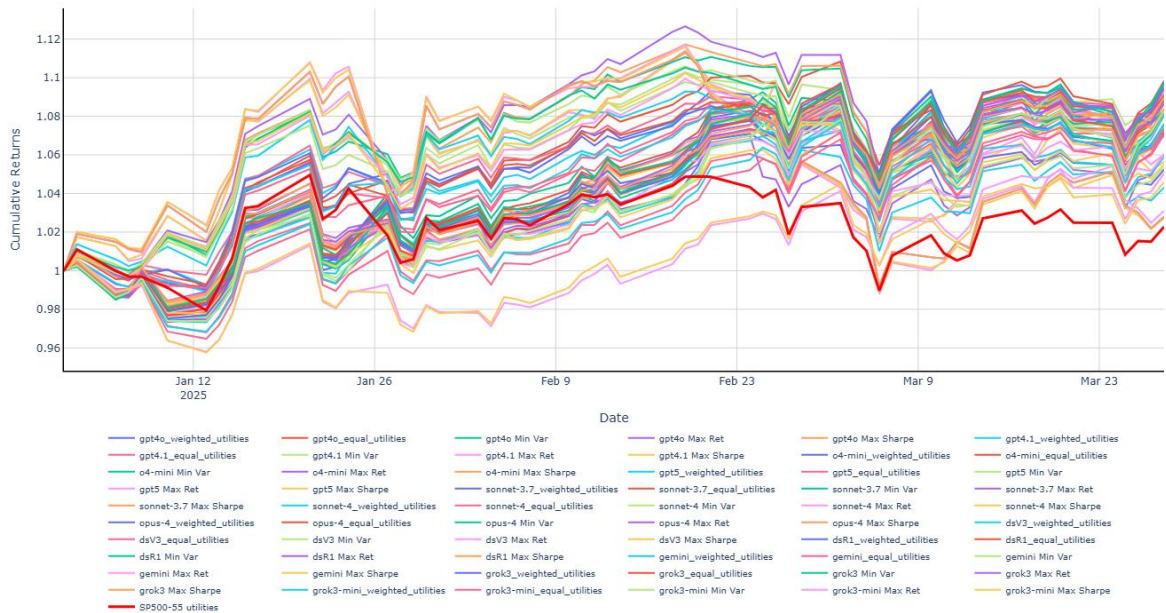

Figure 7: Out-of-sample cumulative returns of portfolios composed by LLMs for "Utilities" sector (January - March 2025)

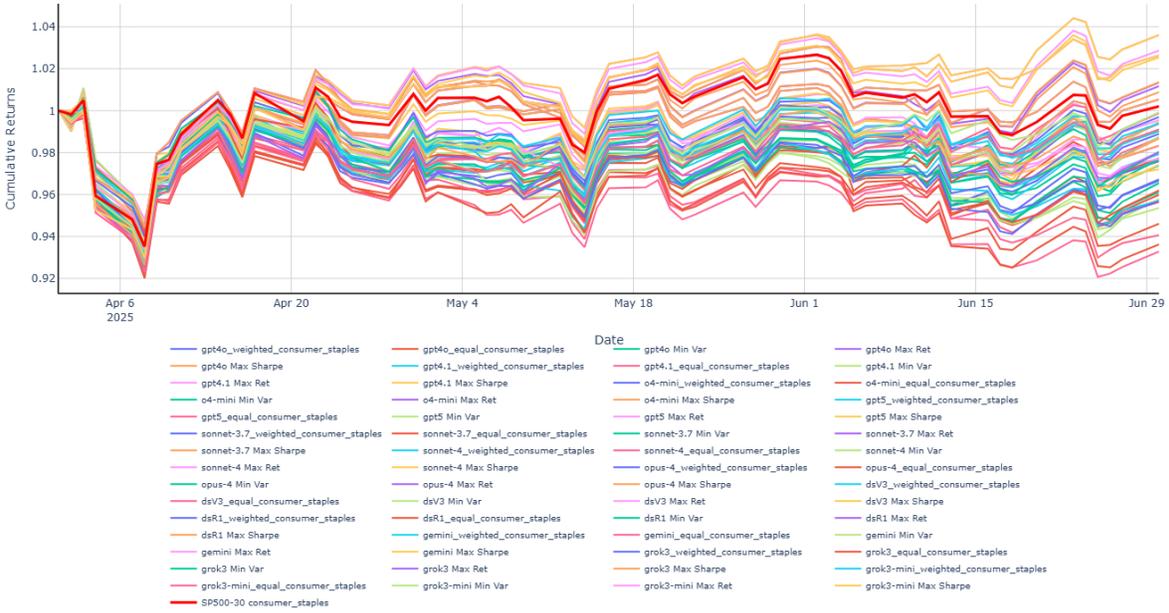

Figure 8: Out-of-sample cumulative returns of portfolios composed by LLMs for "Consumer Staples" sector (April - June 2025)

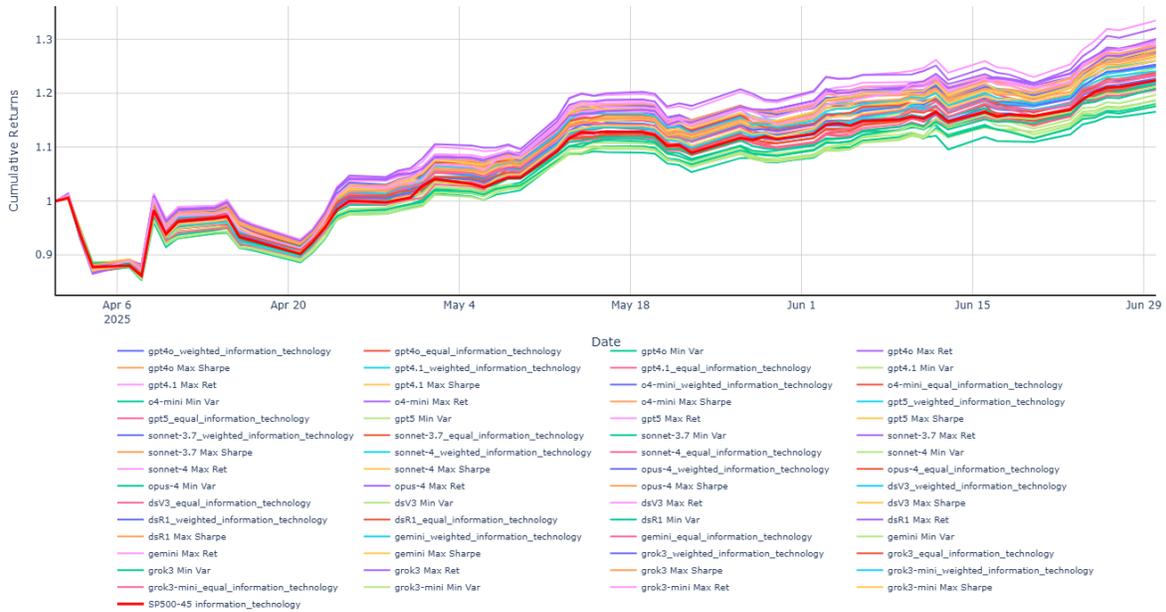

Figure 9: Out-of-sample cumulative returns of portfolios composed by LLMs for "Information Technology" sector (April - June 2025)

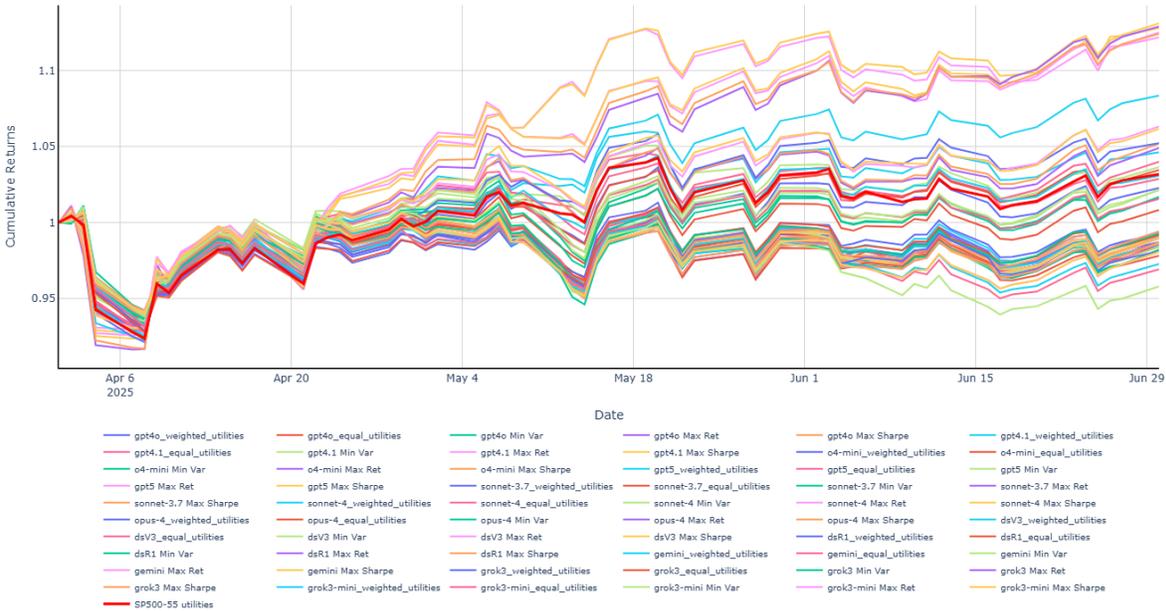

Figure 10: Out-of-sample cumulative returns of portfolios composed by LLMs for "Utilities" sector (April - June 2025)

During the first out-of-sample period (January-March 2025), markets were relatively stable, and most sectors exhibited moderate volatility around baseline levels. In contrast, the second period (April-June 2025) was marked by sharp disruptions and a subsequent recovery. The turbulence began with President Trump's tariff announcement on April 2, which triggered the largest single-day decline in the S&P 500 since 2020 and raised fears of a global trade war [20]. As Figures 8 - 10 show, cumulative returns dropped steeply in early April falling to around 0.85 - 0.90 in several sectors before rebounding as markets stabilized.

These contrasting environments help illustrate how LLM-generated portfolios behave under different market conditions. In stable markets, many models produced portfolios that outperformed their sector benchmarks, whereas performance weakened once volatility rose sharply.

The figures highlight clear cross-sector differences. In some sectors, LLM-weighted portfolios performed best suggesting that the models' stock-picking and weighting decisions were effective. In others, equally weighted or optimization-based portfolios (minimum variance, maximum return, or maximum Sharpe ratio) outperformed the sector index, indicating that while LLM stock selection was valuable, their weight assignments were less optimal. In several cases, however, the sector index itself outperformed all LLM portfolios, underscoring the models' limitations in adapting to certain market regimes.

For example, in the Information Technology sector (Figures 6 and 9), nearly every LLM achieved at least one portfolio that outperformed the benchmark, demonstrating consistent strength in this area. By contrast, in Consumer Staples (Figures 5 and 8), almost all portfolio curves remained below the index throughout the periods, signaling persistent underperformance. Results in sectors like Utilities were mixed, requiring further examination using summary metrics.

To consolidate results across models and sectors, we calculated the mean out-of-sample cumulative return for every portfolio and compared it to its corresponding sector index. A portfolio was considered to outperform its benchmark if its mean cumulative return exceeded that of the sector index. The results are summarized in Table 4 (January-March 2025) and Table 5 (April-June 2025).

The tables use the following notation:

- Green cross pattern: LLM-weighted portfolio outperforms the sector index (the LLM performs well in both stock selection and weighting).
- Yellow lined pattern: At least one of the other portfolios (equally weighted or optimization-based: minimum variance, maximum expected return, or maximum Sharpe ratio) outperforms the sector index (the LLM is strong in stock selection but weaker in weighting).
- Red dotted pattern: The sector index outperforms all portfolios.

From Table 4, two sectors, Industrials and Consumer Staples, stand out as consistent underperformers, with most models failing to build portfolios that beat their respective indices even after optimization. In the later period (Table 5, April-June 2025), the weakest sectors shift slightly to Consumer Discretionary, Consumer Staples, Communication Services, and Utilities, with "Consumer Staples" notably remaining the only sector that underperformed in both time windows.

This temporal shift suggests that LLM portfolio performance is highly dependent on both market regime and sector characteristics. No single model consistently dominated across all sectors or periods; rather, results tended to cluster within sectors, implying that market conditions and sector volatility played a stronger role than model architecture alone.

Conversely, several sectors, especially Energy, Financials, and Information Technology, showed consistent outperformance across both periods. None of these sectors displayed red cells in either table, suggesting they are more conducive to LLM-based stock-picking strategies.

It is also worth noting that while combining LLM-selected stocks with classical mean-variance optimization (to produce minimum variance, maximum expected return, or maximum Sharpe portfolios) improved results in some cases, optimization alone did not guarantee outperformance. Otherwise, there would be no red cells in Tables 4 and 5. This indicates that even when supported by quantitative methods, the underlying quality of LLM-generated stock universes (i.e., selection) remains an important performance driver.

Overall, the evidence from this section shows that LLMs can construct competitive portfolios during stable periods, but their performance deteriorates in times of market stress. Consumer Staples emerges as the least successful sector for LLM strategies, while Information Technology, Financials, and Energy stand out as the most robust. These findings highlight both the promise and the current limitations of LLM-based portfolio construction, effective under normal market regimes but challenged by rapid shifts in volatility and economic uncertainty.

| Reason/ LLM Model | GPT-4o | GPT-4.1 | o4-mini | GPT-5 | Claude Sonnet 3.7 | Claude Sonnet 4 | Claude Opus 4 | DeepSeek -V3 | DeepSeek -R1 | Gemini 2.5 Pro | Grok 3 | Grok 3 Mini |
|---|---|---|---|---|---|---|---|---|---|---|---|---|
| Energy | 🟨 | 🟩 | 🟨 | 🟩 | 🟩 | 🟩 | 🟩 | 🟩 | 🟨 | 🟩 | 🟩 | 🟩 |
| Materials | 🟨 | 🟥 | 🟥 | 🟨 | 🟩 | 🟥 | 🟩 | 🟥 | 🟨 | 🟩 | 🟩 | 🟩 |
| **Industrials** | 🟨 | 🟩 | 🟥 | 🟥 | 🟨 | 🟥 | 🟥 | 🟨 | 🟥 | 🟨 | 🟥 | 🟥 |
| Consumer Discretionary | 🟩 | 🟩 | 🟩 | 🟩 | 🟩 | 🟩 | 🟩 | 🟩 | 🟩 | 🟩 | 🟩 | 🟩 |
| **Consumer Staples** | 🟨 | 🟥 | 🟨 | 🟥 | 🟥 | 🟥 | 🟥 | 🟥 | 🟥 | 🟥 | 🟨 | 🟨 |
| Health Care | 🟥 | 🟨 | 🟩 | 🟩 | 🟩 | 🟩 | 🟩 | 🟩 | 🟥 | 🟩 | 🟥 | 🟩 |
| Financials | 🟨 | 🟩 | 🟨 | 🟩 | 🟩 | 🟩 | 🟨 | 🟩 | 🟨 | 🟨 | 🟨 | 🟨 |
| Information Technology | 🟩 | 🟩 | 🟩 | 🟩 | 🟨 | 🟩 | 🟩 | 🟩 | 🟩 | 🟩 | 🟩 | 🟩 |
| Communication Services | 🟩 | 🟩 | 🟩 | 🟩 | 🟩 | 🟩 | 🟩 | 🟩 | 🟩 | 🟩 | 🟩 | 🟩 |
| Utilities | 🟩 | 🟩 | 🟩 | 🟩 | 🟩 | 🟩 | 🟩 | 🟩 | 🟩 | 🟩 | 🟩 | 🟩 |
| Real Estate | 🟩 | 🟩 | 🟥 | 🟩 | 🟨 | 🟩 | 🟩 | 🟨 | 🟩 | 🟥 | 🟥 | 🟥 |

Table 4: Portfolio performance based on out-of-sample cumulative returns (January-March 2025)

## 6 Portfolio Performance Analysis

This section examines why portfolio performance varied across sectors and time periods by testing several hypotheses related to diversification, dimensionality, volatility, and model consistency. Most structural factors, such as the number of principal components, Herfindahl-Hirschman Index, effective rank, and variations in stock selection or weighting, showed no significant correlation with portfolio outcomes.

The only meaningful relationship observed was with relative volatility, defined as portfolio volatility compared to its sector index. Underperforming portfolios generally exhibited lower relative volatility, implying that they took on less risk and therefore had less potential to exceed benchmark returns. This supports the classic risk-return trade-off, indicating that overly conservative LLM portfolios are less likely to outperform.

| Reason/ LLM Model | GPT-4o | GPT-4.1 | o4-mini | GPT-5 | Claude Sonnet 3.7 | Claude Sonnet 4 | Claude Opus 4 | DeepSeek -V3 | DeepSeek -R1 | Gemini 2.5 Pro | Grok 3 | Grok 3 Mini |
|---|---|---|---|---|---|---|---|---|---|---|---|---|
| Energy | 🟢 | 🟢 | 🟢 | 🟡 | 🟢 | 🟡 | 🟢 | 🟢 | 🟢 | 🟡 | 🟢 | 🟢 |
| Materials | 🟢 | 🟢 | 🟡 | 🟢 | 🟢 | 🟡 | 🟢 | 🟡 | 🟡 | 🟡 | 🟡 | 🟡 |
| Industrials | 🔴 | 🟡 | 🔴 | 🟢 | 🟡 | 🟡 | 🟡 | 🟡 | 🟡 | 🟡 | 🟡 | 🟡 |
| Consumer Discretionary | 🟡 | 🔴 | 🔴 | 🟡 | 🟡 | 🔴 | 🔴 | 🔴 | 🔴 | 🔴 | 🔴 | 🔴 |
| Consumer Staples | 🟡 | 🔴 | 🔴 | 🔴 | 🔴 | 🔴 | 🔴 | 🔴 | 🔴 | 🔴 | 🟡 | 🟡 |
| Health Care | 🟡 | 🟡 | 🟢 | 🟢 | 🟡 | 🟢 | 🟢 | 🟡 | 🟡 | 🟢 | 🟢 | 🟢 |
| Financials | 🟢 | 🟢 | 🟢 | 🟢 | 🟢 | 🟢 | 🟢 | 🟢 | 🟢 | 🟢 | 🟢 | 🟢 |
| Information Technology | 🟢 | 🟢 | 🟢 | 🟢 | 🟢 | 🟢 | 🟢 | 🟢 | 🟢 | 🟢 | 🟢 | 🟢 |
| Communication Services | 🔴 | 🔴 | 🔴 | 🔴 | 🔴 | 🔴 | 🔴 | 🔴 | 🔴 | 🔴 | 🔴 | 🔴 |
| Utilities | 🔴 | 🔴 | 🟢 | 🟡 | 🟢 | 🟢 | 🟢 | 🟢 | 🔴 | 🟢 | 🔴 | 🔴 |
| Real Estate | 🟢 | 🟢 | 🟢 | 🟢 | 🟢 | 🟢 | 🟢 | 🟢 | 🟢 | 🟢 | 🟡 | 🟡 |

Table 5: Portfolio performance based on out-of-sample cumulative returns (April-June 2025)

Overall, the analysis suggests that sector performance differences arise mainly from risk exposure and market volatility, rather than from internal portfolio composition. LLMs tend to perform best when their constructed portfolios maintain a balanced level of risk relative to the sector benchmark.

## 6.1 Principal Component Analysis

This subsection tests whether there is a relationship between a portfolio's performance and the number of principal components required to explain 95 % of its total variance. In simple terms, we explore whether portfolios with a more complex risk structure (i.e., requiring more components) or a more concentrated one (fewer components) tend to perform better out of sample. This analysis aims to determine whether the dimensionality of portfolio risk and return, as captured by principal component analysis (PCA), has any predictive value for subsequent portfolio quality.

PCA was applied to the out-of-sample returns of each portfolio constructed by the twelve LLMs across the eleven S&P 500 sectors. For every portfolio, we calculated the cumulative explained variance of its principal components and recorded the number of components needed to reach the 95% threshold. Figure 11 illustrates this process for the Information Technology sector, where the red dashed line marks the 95% variance level.

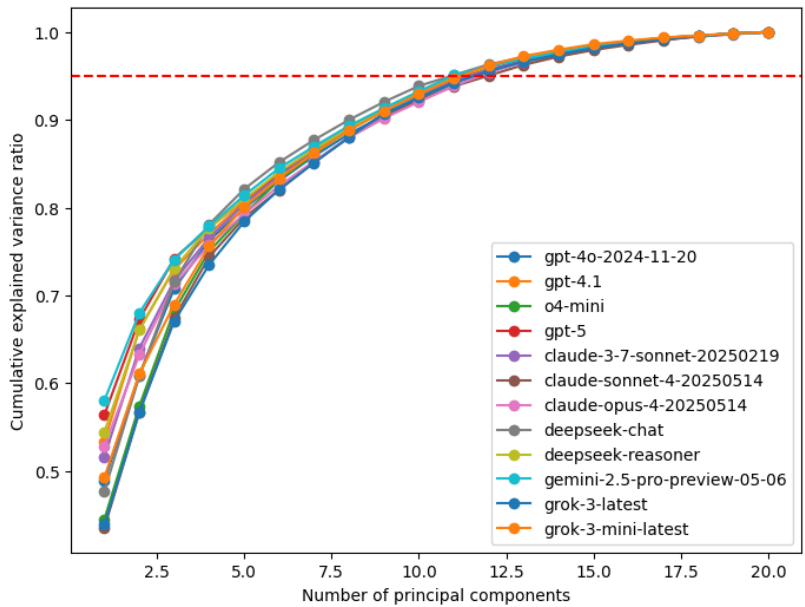

Figure 11: PCA for "Information Technology" sector for portfolios obtained by twelve LLMs

The results reveal that the number of components varies across sectors but shows no consistent pattern linked to performance. The Consumer Staples sector, which consistently underperformed in both out-of-sample periods, required about 9-11 components to explain 95 % of its variance. In comparison, Industrials, another weak performer during the January-March 2025 period, required 11-12 components, while sectors that struggled in the April-June 2025 period (Consumer Discretionary, Communication Services, and Utilities) needed 11-13, 12, and 5-10 components, respectively.

Well-performing sectors showed a similarly wide range: Energy around 10, Materials about 12-13, Financials roughly 10-11, Information Technology 11-12, Health Care 12-14, and Real Estate 10-11 components. This dispersion demonstrates that strong and weak portfolios alike span similar levels of variance complexity.

In summary, no statistically significant relationship was found between the number of principal components explaining 95% of a portfolio's variance and its out-of-sample performance. In other words, portfolios with higher or lower dimensionality do not systematically perform better or worse. This suggests that the underlying structure of return variance, as captured by PCA, does not explain why some LLM-based portfolios outperform while others lag behind.

### 6.2 Herfindahl–Hirschman Index

The Herfindahl-Hirschman Index (HHI) is a standard measure of concentration used in economics and finance [21]. In the context of portfolio analysis, it serves as an indicator of diversification: higher HHI values imply a more concentrated portfolio (greater weight in fewer stocks), while lower values indicate broader diversification. The index is calculated as the sum of the squared portfolio weights:

$$HHI = \sum_{i=0}^{n}(w_i)^2$$

where $w_i$ is the weight of the $i$-th stock in the portfolio. For example, for the portfolio of 20 stocks with equal weights assigned to each stock (1/20 = 0.05), the HHI index will be equal to 0.05. In the case of the portfolios that were composed previously, this measure is just the sum of squares of weights in the portfolio. The HHI index is calculated for each of the portfolios.

In this subsection, we test whether there is a relationship between the level of portfolio concentration (HHI) and out-of-sample performance. If concentration or diversification materially affects returns, sectors with higher or lower HHI values should exhibit consistent performance differences.

The HHI was computed for all portfolios generated by the twelve LLMs across the eleven S&P 500 sectors. The results reveal no consistent trend between diversification and performance. In both outperforming and underperforming sectors, the HHI values fall within comparable ranges. The following insert provides an example from the Consumer Staples sector, which was one of the weakest performers across both out-of-sample periods. The index values for this sector range narrowly across models:

> *Claude Sonnet 3.7 (0.0693), Claude Opus 4 (0.0697), Claude Sonnet 4 (0.0802), DeepSeek-V3 (0.0694), DeepSeek-R1 (0.0673), Gemini (0.0654), GPT-4.1 (0.0697), GPT-4o (0.0613), GPT-5 (0.0600), Grok 3 (0.0621), Grok 3 Mini (0.0575), and o4-mini (0.0565).*

These values indicate only minor variation in diversification levels among models, yet the sector's performance remained consistently poor. Similar patterns appear across other sectors: strong performers such as Energy and Information Technology did not exhibit systematically higher or lower HHI values.

In summary, the findings show no statistically significant relationship between the Herfindahl-Hirschman Index and portfolio performance. Diversification, as measured by HHI, does not explain the performance differences among LLM-generated portfolios. Even sectors with nearly identical concentration levels displayed opposite performance outcomes. Therefore, the hypothesis that portfolio concentration correlates with out-of-sample returns is rejected.

## 6.3 Matrix Effective Rank

The effective rank of a matrix measures the true dimensionality of its data structure, that is, how many components meaningfully contribute to its variability. In the context of portfolio analysis, this metric helps assess how diversified or interdependent asset returns are. A higher effective rank indicates a portfolio with more independent sources of risk, while a lower effective rank suggests higher correlation among assets.

Formally, the effective rank is derived from the Shannon entropy of the eigenvalues of a covariance matrix [22]. Let $A$ denote the covariance matrix of portfolio returns, and let $\lambda_1, \ldots, \lambda_Q$ represent its eigenvalues. After standardizing the eigenvalues so that $p_i = \lambda_i / \sum_{k=1}^{Q} \lambda_k$, the Shannon entropy $H$ is computed as:

$$H(p_1, p_2, \ldots, p_Q) = -\sum_{k=1}^{Q} p_k \log p_k$$

and the effective rank is then defined as

$$erank(A) = e^{H(p_1, p_2, \ldots, p_Q)}.$$

This measure captures how evenly distributed the eigenvalues are: a portfolio dominated by a few principal sources of variance will have a lower effective rank, while one with many relatively equal eigenvalues will have a higher rank.

In this subsection, we test whether portfolio performance is correlated with the effective rank, that is, whether portfolios characterized by more diversified return structures tend to perform better out of sample.

The effective rank was calculated for each portfolio across the twelve LLMs and eleven sectors using the out-of-sample return covariance matrices. The results reveal no clear or systematic relationship between the effective rank and portfolio performance. Both the underperforming and outperforming sectors exhibited a similar range of effective rank values. Sectors such as Consumer Staples and Industrials, which showed weak results, did not have distinctly lower ranks compared with consistently strong performers like Energy or Information Technology.

This finding suggests that the effective dimensionality of the covariance structure (and by extension, the degree of statistical diversification) does not meaningfully predict portfolio success in this study. Portfolios with high effective ranks did not systematically outperform or underperform those with lower ranks.

In summary, the hypothesis that portfolio performance depends on the effective rank of the return covariance matrix is rejected. The dimensional structure of return variance does not appear to drive the differences observed in LLM-generated portfolio outcomes (see [22]).

### 6.4 Number of Stocks in the Universe before Portfolio Selection

Before constructing the final portfolios, each large language model (LLM) was asked to generate a universe of candidate stocks for every S&P 500 sector. Because model outputs vary across runs, this initial universe typically included more than 20 stocks, even though the final portfolios were restricted to the 20 most frequently selected names. The size of this pre-selection universe depends partly on the total number of companies in a given sector (smaller sectors naturally yield smaller candidate sets) yet it also reflects how broadly each LLM interprets its prompt.

This subsection tests two related hypotheses. First, we examine whether the size of the initial stock universe correlates with portfolio performance. Second, we investigate whether the number of "false" stocks (companies suggested by an LLM that do not belong to the requested sector) affects outcomes. The intuition behind these tests is that a model's weaker familiarity with a particular sector might manifest through a larger or less accurate initial universe, potentially leading to poorer final portfolios.

Each LLM's proposed universe was validated against the official S&P 500 sector constituents, as described in Section 3. Any stock not belonging to the correct sector index was labeled a "false" entry and excluded from further analysis. Table 6 summarizes the average universe sizes and false-stock counts by sector.

| LLM Model / Sector | GPT-4o | GPT-4.1 | o4-mini | GPT-5 | Claude Sonnet 3.7 | Claude Sonnet 4 | Claude Opus 4 | DeepSeek-V3 | DeepSeek-R1 | Gemini 2.5 Pro | Grok 3 | Grok 3 Mini | # companies in sector index |
|---|---|---|---|---|---|---|---|---|---|---|---|---|---|
| Energy | 31/11 | 31/10 | 33/12 | 25/4 | 41/20 | 38/17 | 84/64 | 37/16 | 34/13 | 27/6 | 32/11 | 26/5 | 23 |
| Materials | 35/10 | 31/6 | 36/12 | 33/8 | 37/12 | 37/12 | 35/10 | 34/10 | 38/13 | 31/7 | 33/8 | 30/7 | 26 |
| Industrials | 43/6 | 56/5 | 48/8 | 48/2 | 55/4 | 46/3 | 54/5 | 40/4 | 46/6 | 40/1 | 35/3 | 38/3 | 78 |
| Consumer Discretionary | 41/8 | 40/8 | 45/15 | 44/6 | 62/21 | 55/20 | 43/12 | 43/10 | 43/6 | 34/4 | 33/3 | 44/9 | 51 |
| Consumer Staples | 34/0 | 37/2 | 40/5 | 37/3 | 36/3 | 33/4 | 36/2 | 30/0 | 35/2 | 33/1 | 30/2 | 34/0 | 38 |
| Health Care | 37/2 | 42/1 | 42/4 | 55/4 | 42/3 | 44/6 | 44/8 | 38/6 | 49/4 | 38/3 | 28/1 | 34/3 | 60 |
| Financials | 42/6 | 45/7 | 56/8 | 57/3 | 50/6 | 48/3 | 46/7 | 32/1 | 55/7 | 41/2 | 36/1 | 36/0 | 73 |
| Information Technology | 43/13 | 40/9 | 53/16 | 53/4 | 48/9 | 52/15 | 52/14 | 44/17 | 62/23 | 45/6 | 39/7 | 33/5 | 69 |
| Communication Services | 29/9 | 36/16 | 37/15 | 29/6 | 36/16 | 55/33 | 34/14 | 39/19 | 43/23 | 35/15 | 29/9 | 30/8 | 23 |
| Utilities | 33/3 | 35/4 | 38/8 | 32/1 | 39/9 | 34/3 | 34/4 | 35/6 | 34/6 | 31/5 | 29/2 | 29/3 | 31 |
| Real Estate | 39/10 | 43/14 | 42/14 | 37/6 | 44/16 | 47/17 | 45/16 | 35/7 | 30/3 | 40/15 | 37/9 | 33/5 | 31 |

Table 6: Number of stocks in the selection universe and number of stocks in the selection universe not included in the sector index – by sector

The results show no meaningful relationship between the number of candidate stocks or the number of false stocks and portfolio performance. Sectors that underperformed, such as Consumer Staples, did not have unusually large candidate universes or high false-stock counts. In fact, Consumer Staples, the only sector that consistently underperformed in both out-of-sample periods, had one of the lowest average numbers of false stocks (2.0) among all sectors.

These findings indicate that LLM familiarity or data coverage at the sector level does not directly influence portfolio outcomes. Larger or less precise stock universes do not necessarily lead to weaker performance, and the number of excluded "false" stocks appears to have little effect. Consequently, both hypotheses are rejected.

**6.5 Volatility of Portfolios and Index**

Volatility is one of the most fundamental indicators of portfolio risk and stability. In this subsection, we test several hypotheses regarding how volatility (both absolute and relative) relates to the out-of-sample performance of the LLM-constructed portfolios. Specifically, we investigate whether:

- Poor-performing portfolios are concentrated in sectors that exhibit higher market volatility during the out-of-sample period;
- Portfolio-level volatility correlates with performance across sectors; and
- The relative volatility of a portfolio, measured as the ratio of its volatility to that of its corresponding sector index, provides insights into risk-adjusted performance.

The underlying intuition is that if a sector index experiences high volatility, it may be more difficult for LLMs to build portfolios that consistently outperform it. Likewise, a portfolio's relative volatility could indicate whether it is taking too little or too much risk to achieve excess returns.

To test these hypotheses, we calculated the out-of-sample volatilities for each sector index and for every LLM-generated portfolio across both evaluation periods. The relative volatility metric was then obtained by dividing the portfolio's volatility by that of the sector index. Tables 7 and 8 summarize these relative volatilities for the January-March 2025 and April-June 2025 periods, respectively, with underperforming sectors highlighted in grey.

| LLM Model / Sector | GPT-4o | GPT-4.1 | o4-mini | GPT-5 | Claude Sonnet 3.7 | Claude Sonnet 4 | Claude Opus 4 | DeepSeek-V3 | DeepSeek-R1 | Gemini 2.5 Pro | Grok 3 | Grok 3 Mini |
|---|---|---|---|---|---|---|---|---|---|---|---|---|
| Energy | 1.067 | 1.042 | 1.083 | 1.081 | 1.062 | 1.048 | 1.056 | 1.067 | 1.111 | 1.050 | 1.039 | 1.035 |
| Materials | 1.059 | 1.003 | 1.128 | 1.032 | 1.047 | 1.077 | 1.026 | 0.996 | 1.014 | 1.048 | 1.005 | 1.052 |
| Industrials | 0.829 | 0.829 | 0.824 | 0.996 | 0.984 | 0.882 | 0.832 | 0.874 | 0.942 | 0.721 | 0.877 | 0.762 |
| Consumer Discretionary | 0.649 | 0.614 | 0.652 | 0.548 | 0.680 | 0.711 | 0.670 | 0.699 | 0.626 | 0.668 | 0.663 | 0.558 |
| Consumer Staples | 0.870 | 0.870 | 0.863 | 0.951 | 0.848 | 0.905 | 0.843 | 1.163 | 0.748 | 0.946 | 0.774 | 0.984 |
| Health Care | 1.036 | 1.093 | 1.307 | 0.953 | 1.289 | 1.106 | 1.380 | 0.988 | 1.055 | 0.971 | 1.605 | 1.449 |
| Financials | 1.111 | 1.070 | 1.128 | 0.901 | 1.036 | 1.066 | 1.204 | 1.331 | 1.219 | 1.030 | 1.393 | 1.207 |
| Information Technology | 0.909 | 1.022 | 0.858 | 1.126 | 1.081 | 0.956 | 0.928 | 0.981 | 1.117 | 1.109 | 0.887 | 0.922 |
| Communication Services | 0.595 | 0.583 | 0.585 | 0.621 | 0.563 | 0.647 | 0.654 | 0.654 | 0.664 | 0.713 | 0.590 | 0.602 |
| Utilities | 1.919 | 1.966 | 2.077 | 1.523 | 1.690 | 1.527 | 1.409 | 1.900 | 1.822 | 1.440 | 2.128 | 1.896 |
| Real Estate | 1.010 | 1.090 | 0.875 | 1.114 | 1.002 | 1.088 | 1.081 | 1.017 | 1.148 | 0.927 | 1.011 | 0.918 |

Table 7: Relative portfolio volatility across sectors and models (January-March 2025)

The results reveal that neither absolute sector volatility nor portfolio volatility alone is strongly correlated with performance. However, a clear trend emerges when considering their ratio. In sectors where LLM-constructed portfolios underperformed their benchmarks, the relative volatilities were generally lower compared with those of outperforming sectors. In other words, weaker portfolios tended to exhibit more conservative risk profiles relative to their indices.

This finding aligns with the classical risk–return trade-off principle in finance: portfolios that assume less risk than their benchmark may also have limited potential to generate superior returns. While lower volatility reduces downside exposure, it can also restrict upside gains, making it harder to outperform a volatile benchmark.

| LLM Model / Sector | GPT-4o | GPT-4.1 | o4-mini | GPT-5 | Claude Sonnet 3.7 | Claude Sonnet 4 | Claude Opus 4 | DeepSeek -V3 | DeepSeek -R1 | Gemini 2.5 Pro | Grok 3 | Grok 3 Mini |
|---|---|---|---|---|---|---|---|---|---|---|---|---|
| Energy | 1.341 | 1.360 | 1.397 | 1.293 | 1.289 | 1.222 | 1.317 | 1.361 | 1.347 | 1.318 | 1.320 | 1.307 |
| Materials | 1.211 | 1.231 | 1.123 | 1.248 | 1.257 | 1.177 | 1.213 | 1.138 | 1.212 | 1.162 | 1.190 | 1.073 |
| Industrials | 0.897 | 0.941 | 0.839 | 1.202 | 1.023 | 0.953 | 0.929 | 0.947 | 1.015 | 0.916 | 0.966 | 0.907 |
| Consumer Discretionary | 0.837 | 0.735 | 0.820 | 0.630 | 0.809 | 0.794 | 0.794 | 0.817 | 0.729 | 0.732 | 0.799 | 0.711 |
| Consumer Staples | 0.821 | 0.790 | 0.809 | 0.792 | 0.841 | 0.823 | 0.915 | 0.870 | 1.123 | 0.862 | 0.841 | 0.789 |
| Health Care | 1.105 | 1.278 | 1.128 | 0.894 | 0.970 | 1.309 | 0.885 | 1.457 | 1.203 | 0.962 | 1.115 | 0.884 |
| Financials | 1.306 | 1.254 | 1.503 | 1.063 | 1.053 | 1.280 | 1.460 | 1.502 | 1.499 | 1.167 | 1.637 | 1.382 |
| Information Technology | 1.008 | 1.060 | 1.030 | 1.151 | 1.097 | 1.061 | 1.073 | 0.967 | 1.147 | 1.104 | 0.983 | 1.072 |
| Communication Services | 0.828 | 0.798 | 0.844 | 0.805 | 0.813 | 0.902 | 0.899 | 0.904 | 0.856 | 0.869 | 0.824 | 0.896 |
| Utilities | 0.573 | 0.576 | 0.567 | 1.209 | 0.890 | 1.555 | 1.294 | 0.574 | 0.616 | 1.003 | 0.554 | 0.594 |
| Real Estate | 1.093 | 1.063 | 1.253 | 1.055 | 1.043 | 1.157 | 1.109 | 1.111 | 1.049 | 1.028 | 1.173 | 1.034 |

Table 8: Relative portfolio volatility across sectors and models (April-June 2025)

Therefore, the hypothesis regarding relative volatility holds: underperforming portfolios tend to be those with lower relative risk compared to their sector indices. This suggests that the magnitude of risk-taking, rather than raw volatility itself, plays a significant role in explaining LLM portfolio outcomes (see Tables 7 and 8).

**6.6 Weight Volatilities before Final Portfolio Construction**

During the portfolio construction process, each large language model (LLM) was prompted multiple times to assign weights to the selected stocks within a given sector. Because LLMs are stochastic by design, their responses naturally vary from run to run. To capture this variability, the weight-assignment prompt was executed five times per sector, and the average normalized weight of each stock was used to form the final portfolio.

This repeated sampling allows us to analyze the volatility of the weights, that is, how much the assigned stock weights fluctuate across different LLM responses. This variability serves as a proxy for the stability and confidence of the model's reasoning: if a model provides consistent weights across runs, it may be operating with higher internal certainty, while large fluctuations could indicate limited information or uncertainty about the sector.

To explore this, we tested whether there is a relationship between weight volatility and portfolio performance. Specifically, we hypothesized that portfolios constructed by models showing higher variability in assigned weights would perform worse out of sample.

Figure 12 illustrates an example of this analysis for the Industrials sector using the GPT-4o model, where weight variability is visualized across the five generated portfolios. The average standard deviation of weights across all 20 stocks was calculated for each model and sector, and the resulting values are summarized in Table 9.

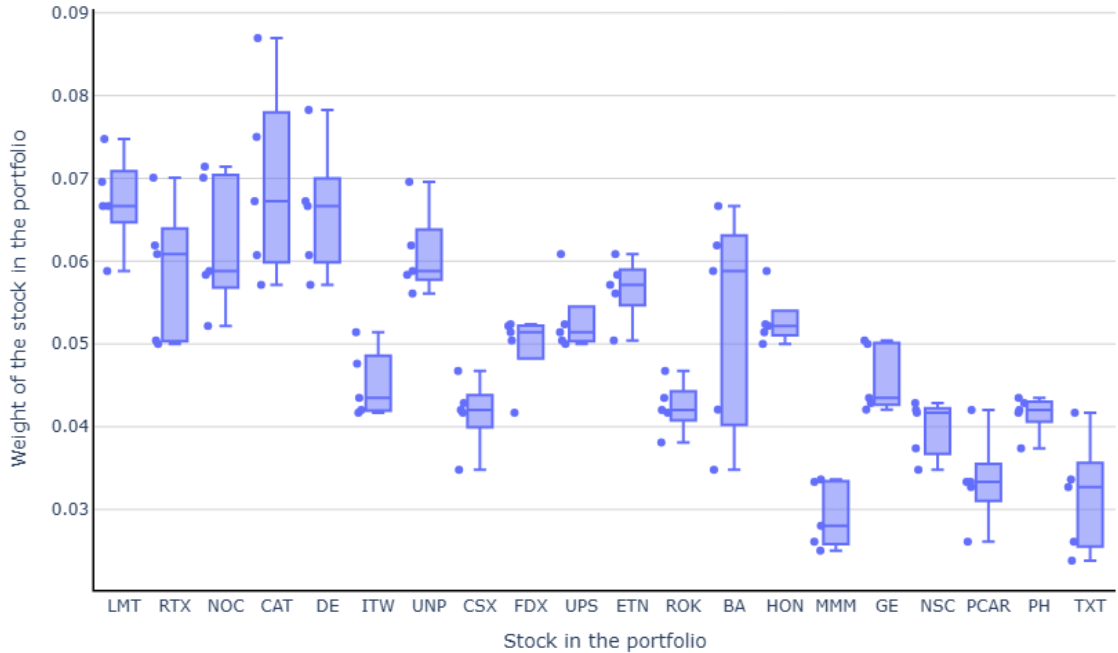

Figure 12: Weight volatility for "Industrials" sector for GPT-4o model

| Sectors / Models | Energy | Materials | Industrials | Consumer Discretionary | Consumer Staples | Health Care | Financials | Information Technology | Communication Services | Utilities | Real Estate |
|---|---|---|---|---|---|---|---|---|---|---|---|
| GPT-4o | 0.0057 | 0.0062 | 0.0052 | 0.0051 | 0.0080 | 0.0069 | 0.0055 | 0.0062 | 0.0055 | 0.0041 | 0.0054 |
| GPT-4.1 | 0.0038 | 0.0054 | 0.0050 | 0.0067 | 0.0053 | 0.0067 | 0.0069 | 0.0074 | 0.0078 | 0.0024 | 0.0085 |
| o4-mini | 0.0149 | 0.0120 | 0.0098 | 0.0123 | 0.0104 | 0.0108 | 0.0119 | 0.0085 | 0.0108 | 0.0106 | 0.0133 |
| GPT-5 | 0.0061 | 0.0061 | 0.0057 | 0.0064 | 0.0061 | 0.0053 | 0.0057 | 0.0060 | 0.0067 | 0.0074 | 0.0064 |
| Claude 3.7 Sonnet | 0.0043 | 0.0044 | 0.0036 | 0.0044 | 0.0045 | 0.0065 | 0.0014 | 0.0067 | 0.0126 | 0.0021 | 0.0048 |
| Claude 4 Sonnet | 0.0125 | 0.0096 | 0.0028 | 0.0050 | 0.0054 | 0.0066 | 0.0095 | 0.0081 | 0.0088 | 0.0049 | 0.0082 |
| Claude 4 Opus | 0.0107 | 0.0107 | 0.0052 | 0.0064 | 0.0062 | 0.0134 | 0.0078 | 0.0054 | 0.0073 | 0.0119 | 0.0116 |
| DeepSeek-V3 | 0.0063 | 0.0049 | 0.0028 | 0.0083 | 0.0032 | 0.0048 | 0.0079 | 0.0046 | 0.0060 | 0.0019 | 0.0064 |
| DeepSeek-R1 | 0.0115 | 0.0095 | 0.0101 | 0.0079 | 0.0079 | 0.0071 | 0.0123 | 0.0066 | 0.0094 | 0.0110 | 0.0055 |
| Gemini 2.5 Pro | 0.0080 | 0.0118 | 0.0040 | 0.0092 | 0.0074 | 0.0089 | 0.0070 | 0.0076 | 0.0090 | 0.0081 | 0.0068 |
| Grok 3 | 0.0028 | 0.0034 | 0.0033 | 0.0040 | 0.0043 | 0.0045 | 0.0060 | 0.0073 | 0.0051 | 0.0027 | 0.0031 |
| Grok 3 Mini | 0.0170 | 0.0150 | 0.0144 | 0.0110 | 0.0142 | 0.0094 | 0.0092 | 0.0097 | 0.0125 | 0.0118 | 0.0123 |

Table 9: Weight volatility across LLMs and sectors

The results show that weight volatility varies not only between sectors but also across models. Notably, the three reasoning-optimized models (o4-mini, DeepSeek-R1, and Grok 3 Mini) displayed higher average weight volatility than the other LLMs. This may reflect their different reasoning styles, which emphasize deliberate, step-by-step evaluation rather than deterministic output generation. However, when these variations were compared to out-of-sample performance, no consistent relationship emerged.

In other words, models that produced more stable weights did not necessarily deliver stronger portfolio results, and those with higher variability did not systematically underperform. Therefore, the hypothesis that weight volatility correlates with portfolio performance is rejected.

Overall, this analysis suggests that fluctuations in weight assignment reflect the reasoning dynamics of different LLM architectures rather than their ability to generate successful portfolios. While variability in weights may indicate differing reasoning depth or stochastic tendencies, it does not appear to influence the risk-adjusted outcomes of LLM-constructed portfolios (see Figure 12 and Table 9).

### 6.7 Performance of the Sector Index

Another potential explanation for the variation in portfolio outcomes is that some sectors may already be strong performers during the evaluation period, making it inherently more difficult for LLM-generated portfolios to outperform them. Conversely, sectors experiencing weaker market performance might offer greater opportunities for excess returns. To test this, we compared the mean cumulative return of each sector with the relative success of its corresponding LLM portfolios.

The mean cumulative out-of-sample returns for all eleven sector indices were calculated for both evaluation periods. The results are presented in Tables 10 and 11, which list the sectors in ascending order of average return.

| Sector | Mean cumulative out-of-sample return |
| --- | --- |
| Information Technology | 0.961992 |
| Consumer Discretionary | 0.973021 |
| Communication Services | 1.015604 |
| Utilities | 1.022453 |
| **Industrials** | **1.028976** |
| Real Estate | 1.031273 |
| **Consumer Staples** | **1.032195** |
| Energy | 1.037267 |
| Financials | 1.039232 |
| Materials | 1.050402 |
| Health Care | 1.055187 |

Table 10: Out-of-sample cumulative returns of the sector indices (January-March 2025)

| Sector | Mean cumulative out-of-sample return |
|---|---|
| Energy | 0.881505 |
| Health Care | 0.939217 |
| Real Estate | 0.978392 |
| Materials | 0.98324 |
| Financials | 0.993297 |
| **Consumer Staples** | **0.999588** |
| **Utilities** | **1.004957** |
| **Consumer Discretionary** | **1.028861** |
| Industrials | 1.034086 |
| **Communication Services** | **1.046453** |
| Information Technology | 1.068461 |

Table 11: Out-of-sample cumulative returns of the sector indices (April-June 2025)

In the January-March 2025 period (Table 10), the sector indices ranged from a low of 0.961992 for Information Technology to a high of 1.055187 for Health Care. The underperforming sectors identified earlier, Industrials and Consumer Staples, fell in the middle of this range, suggesting that their weak LLM portfolio results were not simply a consequence of strong benchmark performance.

In the April-June 2025 period (Table 11), the ordering of sectors changed significantly due to heightened market volatility. Energy and Health Care recorded the lowest mean cumulative returns (0.881505 and 0.939217, respectively), while Information Technology and Communication Services posted the highest (1.068461 and 1.046453). During this period, the weakest LLM portfolio results appeared in Consumer Discretionary, Consumer Staples, Utilities, and Communication Services, yet these were not the top-performing indices, undermining the notion that strong sector benchmarks are inherently harder to beat.

Overall, the analysis reveals no consistent relationship between a sector's own market performance and the ability of LLMs to outperform it. In some cases, portfolios struggled despite moderate or even weak benchmark returns, while in others they succeeded against stronger sectors.

Therefore, the hypothesis that LLM portfolios underperform simply because their sector indices perform exceptionally well is rejected. The results indicate that sector-level returns alone cannot explain differences in LLM portfolio outcomes. Instead, the findings reinforce the idea that factors such as relative volatility and market stability, rather than the absolute performance of the benchmark, play a more influential role in determining success (see Tables 10 and 11).

## 6.8 Portfolio Performance Comparison based on Sharpe Ratio

To further evaluate the performance of the LLM-constructed portfolios, we compare their out-of-sample Sharpe ratios with those of the corresponding sector indices. The Sharpe ratio captures risk-adjusted performance, measuring excess return per unit of volatility, and is defined as

$$SR_p = \frac{\mathbb{E}(r_p) - r_f}{\sigma_p},$$

where $\mathbb{E}(r_p)$ is the mean portfolio return, the $r_f$ is the risk-free rate of portfolio return, and $\sigma_p$ is the standard deviation of portfolio return (volatility). The risk-free rate is set to 2% (0.02) as per standards.

For each portfolio, the weighted portfolio return was computed using the matrix of out-of-sample stock returns $A$ and the weight vector $w$:

$$A_w = A^T w$$

$A_w$ is taken to compute Sharpe ratios (substituting $\mathbb{E}(r_p)$) for all five portfolio types (LLM-weighted, equally weighted, minimum variance, maximum expected return, and maximum Sharpe ratio) and compared against the sector benchmarks. The results for the sector indices themselves are shown in Tables 12 and 13, while consolidated comparison outcomes appear in Tables 14 and 15.

| Sector name | Sharpe ratio |
|---|---|
| Health Care | 0.120377 |
| Energy | 0.105376 |
| **Consumer Staples** | **0.082759** |
| Real Estate | 0.056000 |
| Materials | 0.055974 |
| Utilities | 0.049539 |
| Financials | 0.048247 |
| **Industrials** | **-0.005294** |
| Communication Services | -0.084248 |
| Information Technology | -0.119906 |
| Consumer Discretionary | -0.145554 |

Table 12: Out-of-sample Sharpe ratios of the sector indices (January-March 2025)

| Sector name | Sharpe ratio |
|---|---|
| Information Technology | 0.134460 |
| **Communication Services** | **0.132602** |
| Industrials | 0.103740 |
| **Consumer Discretionary** | **0.074267** |
| Financials | 0.051017 |
| **Utilities** | **0.039833** |
| Materials | 0.024880 |
| **Consumer Staples** | **0.001913** |
| Real Estate | -0.010021 |
| Health Care | -0.064494 |
| Energy | -0.064933 |

Table 13: Out-of-sample Sharpe ratios of the sector indices (April-June 2025)

In the January-March 2025 period (Table 12), the highest Sharpe ratios were observed for the Health Care (0.120377) and Energy (0.105376) sectors, while Industrials, Communication Services, Information Technology, and Consumer Discretionary displayed negative Sharpe ratios, reflecting weaker risk-adjusted performance. In contrast, during the April-June 2025 period (Table 13), the ranking reversed: Information Technology and Communication Services achieved the strongest risk-adjusted returns (0.134460 and 0.132602), whereas Energy and Health Care dropped to the lowest levels (-0.064933 and -0.064494).

When the LLM portfolios are evaluated against these benchmarks (Tables 14 and 15), a striking temporal difference emerges:

- In the first out-of-sample period (January-March 2025), almost all sectors show green cross-pattern cells, meaning that LLM-weighted portfolios outperformed their sector indices on a risk-adjusted basis. A few yellow cells indicate that optimization-based or equally weighted portfolios also achieved superior Sharpe ratios, but overall, the LLM portfolios demonstrated stronger risk-return efficiency during stable market conditions.
- In contrast, the April-June 2025 results tell a different story. Under heightened volatility, the number of green cells drops sharply, replaced by yellow and red patterns. Sectors such as Energy, Consumer Staples, Health Care, and Utilities show clear underperformance relative to their benchmarks, even after adjusting for risk.

| Reason/ LLM Model | GPT-4o | GPT-4.1 | o4-mini | GPT-5 | Claude Sonnet 3.7 | Claude Sonnet 4 | Claude Opus 4 | DeepSeek -V3 | DeepSeek -R1 | Gemini | Grok 3 | Grok 3 Mini |
|---|---|---|---|---|---|---|---|---|---|---|---|---|
| Energy | 🟩 | 🟩 | 🟩 | 🟩 | 🟩 | 🟩 | 🟩 | 🟩 | 🟩 | 🟩 | 🟩 | 🟩 |
| Materials | 🟩 | 🟩 | 🟩 | 🟩 | 🟩 | 🟩 | 🟩 | 🟩 | 🟩 | 🟩 | 🟩 | 🟩 |
| Industrials | 🟩 | 🟩 | 🟩 | 🟩 | 🟩 | 🟩 | 🟩 | 🟩 | 🟩 | 🟩 | 🟩 | 🟩 |
| Consumer Discretionary | 🟩 | 🟩 | 🟩 | 🟩 | 🟩 | 🟩 | 🟩 | 🟩 | 🟩 | 🟩 | 🟩 | 🟩 |
| Consumer Staples | 🟨 | 🟨 | 🟨 | 🟨 | 🟨 | 🟨 | 🟨 | 🟨 | 🟨 | 🟨 | 🟨 | 🟨 |
| Health Care | 🟩 | 🟩 | 🟩 | 🟩 | 🟩 | 🟩 | 🟩 | 🟩 | 🟩 | 🟩 | 🟩 | 🟩 |
| Financials | 🟩 | 🟩 | 🟩 | 🟩 | 🟩 | 🟩 | 🟩 | 🟩 | 🟩 | 🟩 | 🟩 | 🟩 |
| Information Technology | 🟩 | 🟩 | 🟩 | 🟩 | 🟥 | 🟩 | 🟩 | 🟩 | 🟩 | 🟩 | 🟩 | 🟩 |
| Communication Services | 🟩 | 🟩 | 🟩 | 🟩 | 🟩 | 🟩 | 🟩 | 🟨 | 🟨 | 🟩 | 🟩 | 🟩 |
| Utilities | 🟨 | 🟩 | 🟩 | 🟩 | 🟩 | 🟩 | 🟩 | 🟨 | 🟩 | 🟩 | 🟩 | 🟨 |
| Real Estate | 🟩 | 🟩 | 🟩 | 🟩 | 🟩 | 🟨 | 🟩 | 🟩 | 🟩 | 🟩 | 🟥 | 🟩 |

Table 14: Portfolio performance based on Sharpe ratios (January-March 20205)

This divergence between the two periods underscores a key finding: *LLM-based portfolios perform well in stable market regimes but lose their edge during turbulent conditions.* The increased presence of yellow cells in the second period suggests that while LLMs still identify competitive stock selections, their weighting strategies become less effective when volatility rises.

The observed decline likely stems from information asymmetry in LLM training data: most models have limited exposure to high-volatility or crisis-period market examples, making it difficult for them to adapt to sudden regime changes. As a result, LLM portfolios exhibit weaker risk-adjusted performance when market uncertainty spikes.

In summary, comparing Sharpe ratios across both periods confirms that LLM portfolios achieve superior risk-reward performance under normal conditions but face significant challenges during market stress. This finding reinforces the broader conclusion of the study: LLMs can be valuable tools for portfolio construction and risk management, but their outputs should be complemented by traditional quantitative techniques and human oversight to maintain robustness across varying market environments (see Tables 12-15).

| Reason/LLM Model | GPT-4o | GPT-4.1 | o4-mini | GPT-5 | Claude Sonnet 3.7 | Claude Sonnet 4 | Claude Opus 4 | DeepSeek-V3 | DeepSeek-R1 | Gemini | Grok 3 | Grok 3 Mini |
|---|---|---|---|---|---|---|---|---|---|---|---|---|
| Energy | 🔴 | 🔴 | 🔴 | 🔴 | 🔴 | 🔴 | 🔴 | 🔴 | 🔴 | 🔴 | 🔴 | 🔴 |
| Materials | 🟡 | 🟡 | 🟢 | 🟢 | 🟡 | 🟡 | 🟡 | 🟡 | 🟡 | 🟡 | 🟡 | 🟡 |
| Industrials | 🟡 | 🟢 | 🟡 | 🟡 | 🟡 | 🟡 | 🟡 | 🟡 | 🟡 | 🟡 | 🟡 | 🟢 |
| Consumer Discretionary | 🟡 | 🟡 | 🟢 | 🟢 | 🟡 | 🟡 | 🟡 | 🟡 | 🟡 | 🟡 | 🟡 | 🟡 |
| Consumer Staples | 🟡 | 🔴 | 🟡 | 🟡 | 🟡 | 🔴 | 🔴 | 🟡 | 🔴 | 🟡 | 🔴 | 🟡 |
| Health Care | 🟢 | 🟢 | 🔴 | 🟡 | 🟢 | 🟡 | 🟢 | 🔴 | 🟡 | 🟢 | 🔴 | 🟢 |
| Financials | 🟡 | 🟡 | 🟡 | 🟡 | 🟢 | 🟡 | 🟡 | 🟢 | 🟢 | 🟡 | 🟡 | 🟡 |
| Information Technology | 🟡 | 🟡 | 🟡 | 🟡 | 🟡 | 🟡 | 🟡 | 🟡 | 🟡 | 🟡 | 🟡 | 🟡 |
| Communication Services | 🟡 | 🟡 | 🟡 | 🟡 | 🟡 | 🟡 | 🟡 | 🟡 | 🟡 | 🟡 | 🟡 | 🟡 |
| Utilities | 🔴 | 🔴 | 🔴 | 🟡 | 🟡 | 🟢 | 🟡 | 🔴 | 🔴 | 🔴 | 🔴 | 🔴 |
| Real Estate | 🟡 | 🟡 | 🟢 | 🟢 | 🟡 | 🟡 | 🟡 | 🟡 | 🔴 | 🟡 | 🟡 | 🔴 |

Table 15: Portfolio performance based on Sharpe ratios (Apr – Jun 20205)

## 7 Conclusions and Further Steps

This study set out to evaluate how effectively large language models (LLMs) from leading AI providers (OpenAI, Google, Anthropic, DeepSeek, and xAI) can be applied to the task of quantitative sector-based portfolio construction. By integrating generative reasoning with classical optimization methods, our goal was to determine whether LLMs can meaningfully contribute to stock selection, portfolio weighting, and overall investment performance across different market conditions.

Our research makes several key contributions to the emerging literature on AI-driven finance:

1. We conduct one of the first large-scale, multi-model comparisons of LLMs applied to sector-level investment portfolios, analyzing twelve models across eleven S&P 500 sectors.
2. We evaluate performance over two distinct out-of-sample market regimes, a stable period (January-March 2025) and a volatile period (April-June 2025), to test how model behavior shifts with market dynamics.

3. We combine LLM-based stock selection with traditional mean–variance portfolio optimization, assessing whether this hybrid approach enhances returns and stability.
4. We introduce a series of diagnostic analyses (including PCA, HHI, effective rank, relative volatility, and weight variation) to explain sector-level performance differences and test whether portfolio structure, risk exposure, or data limitations drive results.

The findings provide a detailed view of the role of generative AI in quantitative investment management. Overall, all twelve LLMs demonstrated consistent behavioral patterns across sectors, showing that while their reasoning frameworks differ, their portfolio outcomes tend to align. When evaluated by cumulative returns and Sharpe ratios, LLM portfolios performed competitively during stable markets, often outperforming their sector benchmarks. However, their performance declined notably during volatile periods, suggesting that current models still struggle to adapt to rapidly changing market regimes or to scenarios underrepresented in their training data.

Importantly, our diagnostic analysis identified relative volatility, i.e., the ratio of portfolio to benchmark volatility, as the only measure significantly correlated with performance. Underperforming portfolios tended to exhibit lower relative volatility, meaning they took on less risk and, consequently, generated smaller excess returns. Other structural factors, such as diversification (HHI), principal component dimensionality, or the number of selected stocks, showed no statistical relationship with portfolio success.

These results imply that LLMs currently function best as analytical assistants rather than autonomous investment engines. They can efficiently identify promising stocks, articulate rationales rooted in traditional investment logic, and complement quantitative optimization. However, without mechanisms to dynamically adjust for regime changes or incorporate real-time financial data, their standalone effectiveness remains limited.

Looking forward, several directions emerge for future research:

- Advanced prompt engineering may enhance LLMs' reasoning consistency and improve the precision of stock selection and weighting.
- The inclusion of domain-specific financial models, such as BloombergGPT or future fine-tuned variants, could reveal how specialized training impacts investment accuracy and risk management.
- Exploring dynamic or adaptive portfolio strategies, where LLMs periodically rebalance positions in response to changing market signals, may yield more robust outcomes across cycles.
- Finally, integrating LLMs with real-time data access and hybrid decision frameworks, combining AI insights, quantitative models, and human judgment, offers a promising path toward reliable, explainable, and transparent AI-assisted portfolio management.

In conclusion, this study demonstrates that generative AI can enhance sector-based portfolio construction when used responsibly and in concert with established quantitative techniques. While LLMs are not yet a substitute for human or algorithmic decision-making, they already show significant potential as supportive tools that bridge qualitative reasoning and quantitative rigor. By deepening their integration with financial modeling and ensuring transparency in their logic, future iterations of LLMs could play an increasingly central role in data-driven, adaptive, and interpretable investment management.

# References

1. Li Y, Wang S, Ding H, Chen H (2023). *Large Language Models in Finance: A Survey*. In *Proceedings of the Fourth ACM International Conference on AI in Finance* (pp. 374–382). ICAIF '23. https://doi.org/10.1145/3604237.3626869
2. Zaremba A, Demir E (2023). *ChatGPT: Unlocking the future of NLP in finance.* In *Modern Finance,* 1.1 , (pp. 93–98). https://doi.org/10.61351/mf.v1i1.43
3. Krause D (2023). *Large Language Models and Generative AI in Finance: An Analysis of ChatGPT, Bard, and Bing AI*. http://dx.doi.org/10.2139/ssrn.4511540
4. Romanko O, Narayan A, Kwon RH (2023) ChatGPT-Based Investment Portfolio Selection. In: *Operations Research Forum.* Vol. 4. Springer, article 91. DOI: https://doi.org/10.1007/s43069-023-00277-6
5. Biswas S, Joshi N, Mukhopadhyaya JN (2023) ChatGPT in Investment Decision Making: An Introductory Discussion. https://doi.org/10.13140/RG.2.2.36417.43369
6. Ko H, Lee J (2023). *Can ChatGPT Improve Investment Decision? From a Portfolio Management Perspective*. In *Finance Research Letters* (p. 105433) https://doi.org/10.1016/j.frl.2024.105433
7. Haas C, Gilmore A (2023) *Introducing BloombergGPT, Bloomberg's 50-billion parameter large language model, purpose-built from scratch for Finance*. In *Bloomberg* (2023). https://www.bloomberg.com/company/press/bloomberggpt-50-billion-parameter-llm-tuned-finance/
8. Wu S, Irsoy O, Lu S, Dabravolski V, Dredze M, Gehrmann S, Kambadur P, Rosenberg D, Mann G (2023). *BloombergGPT: A Large Language Model for Finance*. https://doi.org/10.48550/arXiv.2303.17564
9. Fatouros G, Metaxas K, Soldatos J, Kyriazis D (2024). *Can Large Language Models Beat Wall Street? Unveiling the Potential of AI in Stock Selection*. http://dx.doi.org/%2010.2139/ssrn.4693849
10. SP Dow Jones Indices. (2025). *S&P U.S. Indices Methodology*. A division of S&P Global. https://www.spglobal.com/spdji/en/documents/methodologies/methodology-sp-us-indices.pdf
11. S&P Global. (2025). *S&P Global*. https://www.spglobal.com/en/
12. Wikipedia contributors. (n.d.). *List of S&P 500 companies. Wikipedia*. https://en.wikipedia.org/wiki/List_of_S%26P_500_companies
13. TradingView. *S&P 500 - TradingView*. https://www.tradingview.com/markets/indices/quotes-snp/
14. OpenAI. *OpenAI Platform*. https://platform.openai.com/docs/models
15. Jamali L, McMahon L (2025). OpenAI claims GPT-5 model boosts ChatGPT to 'PhD level'. In: BBC News, 8 Aug 2025. https://www.bbc.com/news/articles/c3gw4pe20kpo
16. Google. *Gemini models*. https://ai.google.dev/gemini-api/docs/models
17. Anthropic. (2025). *Introducing Claude 4*. May 2025. Introducing Claude 4 \ Anthropic
18. xAI. Welcome to the xAI documentation. https://docs.x.ai/docs/models. Accessed 16 July 2025.

19. Markowitz HM (1952) *Portfolio Selection. Journal of Finance.* 7 (1), pp. 77-91. http://doi.org/10.2307/2975974
20. Evans B, Melloy J, Singh P (2025) Dow nosedives 1,600 points, S&P 500 and Nasdaq drop the most since 2020 after Trump's tariff onslaught. In: CNBC. https://www.cnbc.com/2025/04/02/stock-market-today-live-updates-trump-tariffs.html. Accessed 20 Jul 2025
21. Rhoades SA (1993). *The Herfindahl-Hirschman Index*. In *Fed. Res. Bull.,* 79 (p. 1888). https://heinonline.org/HOL/LandingPage?handle=hein.journals/fedred79&div=37
22. Roy O, Vetterli M (2007). *The effective rank: A measure of effective dimensionality*. In *2007 15th European Signal Processing Conference* (pp. 606–610). IEEE 2007.